\documentclass[preprintnumbers, floatfix,letterpaper,aps,prd,epsfig,nofootinbib,twocolumn]{revtex4-2}

\usepackage{graphicx}
\usepackage{dcolumn}
\usepackage[colorlinks,linkcolor=blue,citecolor=blue,urlcolor=blue]{hyperref}
\usepackage{bm,graphicx,dcolumn,epstopdf,epsf, latexsym,mathbbol, amssymb,amsmath,color, slashed, mathrsfs,mathcomp,simplewick}
\usepackage[center]{subfigure}
\usepackage{multirow}
\usepackage{makecell}

\usepackage{hyperref}
\hypersetup{
  colorlinks=true,  
  linkcolor=blue,      
  citecolor=blue, 
}

\usepackage[makeroom]{cancel}
\usepackage{amsfonts}
\usepackage[export]{adjustbox}
\usepackage{array}
\usepackage{appendix}
\usepackage{relsize}
\usepackage{subcaption}
\usepackage{enumerate}

\begin{document}
\allowdisplaybreaks

\newcommand{\be}{\begin{equation}}
\newcommand{\ee}{\end{equation}}
\newcommand{\beq}{\begin{equation}}
\newcommand{\eeq}{\end{equation}}
\newcommand{\bea}{\begin{eqnarray}}
\newcommand{\eea}{\end{eqnarray}}
\newcommand{\HH}{{\cal H}}
\newcommand{\bra}[1]{\mbox{$\langle\, #1 \mid$}}
\newcommand{\bbra}[1]{\mbox{$\left\langle\, #1 \right\mid$}}
\newcommand{\ket}[1]{\mbox{$\mid #1\,\rangle$}}
\newcommand{\bket}[1]{\mbox{$\left\mid #1\,\right\rangle$}}
\newcommand{\expec}[1]{\mbox{$\langle\, #1\,\rangle$}}
\newcommand{\expecl}[1]{\mbox{$\left\langle\,\strut\displaystyle{#1}\,\right\rangle$}}
\newcommand{\bsy}{\boldsymbol}
\newcommand{\re}{\Re{\rm e}}
\newcommand{\im}{\Im{\rm m}}
\renewcommand{\a}{\hat a}
\newcommand{\ac}{\hat a^{\dagger}}
\renewcommand{\b}{\hat b}
\newcommand{\bc}{\hat b^\dagger}
\renewcommand{\d}{\mbox{${\rm d}$}}
\newcommand{\lp}{\ell_{\rm p}}
\newcommand{\mpl}{m_{\rm p}}
\newcommand{\gn}{G_{\rm N}}
\newcommand{\rh}{r_{\rm H}}
\newcommand{\rH}{r^{\rm H}}
\newcommand{\Rh}{R_{\rm H}}
\newcommand{\RH}{R^{\rm H}}
\newcommand{\psis}{{\psi}_{\rm S}}
\newcommand{\Hs}{\mathcal{H}_{\rm S}}
\newcommand{\Hh}{\mathcal{H}_{\rm H}}
\newcommand{\psih}{{\psi}_{\rm H}}
\newcommand{\erf}{\mathrm{erf}}
\newcommand{\erfc}{\mathrm{erfc}}
\newcommand{\Ng}{N_{\rm G}}
\newcommand{\eg}{\varepsilon_{\rm G}}
\newcommand{\Nb}{N_{\rm B}}
\newcommand{\NL}{N_{\Lambda}}
\newcommand{\Nh}{N_{\rm H}}

\newcommand{\Rb}{R_{\rm B}}
\newcommand{\mb}{m_{\rm B}}
\newcommand{\Lagr}{\mathcal{L}}

\title{Periodic orbits and their gravitational wave radiations in black hole with dark matter halo}

\author{Sumarna Haroon}
\affiliation{Institute for Theoretical Physics and Cosmology, Zhejiang University of Technology, Hangzhou, 310023, China}
\affiliation{United Center for Gravitational Wave Physics (UCGWP), Zhejiang University of Technology, Hangzhou, 310023, China}
\author{Tao Zhu}
\email{Corresponding author: zhut05@zjut.edu.cn}
\affiliation{Institute for Theoretical Physics and Cosmology,
Zhejiang University of Technology, Hangzhou, 310023, China}
\affiliation{United Center for Gravitational Wave Physics (UCGWP), Zhejiang University of Technology, Hangzhou, 310023, China}
	
\begin{abstract}

This study investigates the gravitational dynamics of a spherically symmetric black hole embedded within a dark matter halo, focusing on the impact of the dark matter halo on time-like particle geodesics. We analyze key orbital parameters, including marginally bound orbits, innermost stable circular orbits, and periodic orbits, demonstrating how the dark matter halo alters their energy and angular momentum compared to an isolated black hole. Furthermore, we calculate the gravitational waveforms emitted by these periodic orbits, providing potential observational signatures for such systems. Our results demonstrate a clear correlation between the gravitational waveforms emitted by a small object orbiting a supermassive black hole and the object's zoom-whirl orbital behavior. Higher zoom numbers correspond to more complex waveform substructures. The presence of a dark matter halo significantly impacts these waveforms. We have conducted a detailed analysis of the frequency spectra of the gravitational waveforms arising from these periodic orbits and evaluated their potential detectability. This research contributes to a deeper understanding of the interplay between black holes and dark matter halos, offering insights into their complex gravitational interactions.

\end{abstract}
	
\maketitle

\section{Introduction}
\renewcommand{\theequation}{1.\arabic{equation}} \setcounter{equation}{0}

The landmark detection of gravitational waves (GWs) from binary black hole mergers by LIGO and Virgo \cite{LIGOScientific:2016aoc, LIGOScientific:2016vbw, LIGOScientific:2016vlm, LIGOScientific:2016emj} has opened a new era for probing strong-field gravity and black hole dynamics. These detections have not only validated predictions of general relativity (GR) in highly nonlinear regimes but have also opened new avenues for studying compact objects and their astrophysical environments. Among the most promising targets for future space-based GW detectors are extreme mass ratio inspirals (EMRIs), where a stellar-mass compact object, such as a black hole or neutron star, gradually spirals into a supermassive black hole (SMBH). Space-based GW detectors like LISA \cite{LISA:2022yao, LISA:2017pwj}, Taiji \cite{Hu:2017mde, Gong:2021gvw}, TianQin \cite{Liu:2020eko, TianQin:2015yph, Gong:2021gvw}, and DECIGO \cite{Musha:2017usi} are ideally suited to detect the long-duration, low-frequency signals emitted by these systems. EMRIs are of particular significance because their gravitational waveforms encode detailed information about the spacetime structure of the central black hole \cite{Hughes:2000ssa, Glampedakis:2005hs, Barausse:2020rsu}. This provides an unparalleled opportunity to probe the Kerr geometry predicted by GR and to search for potential deviations caused by new physics, such as modified gravity or the presence of exotic compact objects \cite{Cardoso:2019rvt, Babak:2017tow}.

A critical feature of EMRI dynamics is the prevalence of periodic orbits$-$bound trajectories in which a particle returns to its initial state after an integer number of radial and angular oscillations \cite{Levin:2008ci,Levin:2008mq}. These orbits, indexed by zoom$-$whirl parameters $(z,w,v)$ \cite{Levin:2008ci,Levin:2008mq, Levin:2009sk}, serve as foundational modes for understanding generic orbital dynamics \cite{Levin:2008mq, Levin:2009sk, Misra:2010pu, Babar:2017gsg}. The framework of periodic orbits has been extensively applied to a variety of black hole spacetimes, including charged black holes \cite{Misra:2010pu}, naked singularities \cite{Babar:2017gsg}, Kerr-Sen black holes \cite{Liu:2018vea}, and Einstein-Lovelock black holes \cite{Lin:2021noq}, Kehagias-Sfetsos black holes \cite{Wei:2019zdf}, quantum-corrected black holes \cite{Deng:2020yfm} and brane-world black holes \cite{Deng:2020hxw}. For further studies of periodic orbits in other black hole spacetimes, see Refs.~\cite{Lin:2023eyd, Chan:2025ocy, Zhou:2020zys, Lin:2022wda, Zhang:2022psr, Lin:2022llz, Habibina:2022ztd, Zhang:2022zox, Wang:2022tfo, Lin:2023rmo, Yao:2023ziq, Tu:2023xab, Yang:2024lmj, Shabbir:2025kqh, Junior:2024tmi, Zhao:2024exh, Jiang:2024cpe, Yang:2024cnd, Meng:2024cnq, Li:2024tld, QiQi:2024dwc, Lu:2025cxx}. Gravitational waveforms from EMRIs inherit distinct signatures of these periodic orbits, particularly the characteristic ``zoom-whirl" phases during the inspiral \cite{Tu:2023xab, Yang:2024lmj}. This has motivated numerous studies on the GWs emitted by periodic orbits in various black hole spacetimes, as well as their potential detectability by future space-based GW observatories \cite{Tu:2023xab, Yang:2024lmj, Shabbir:2025kqh, Junior:2024tmi, Zhao:2024exh, Jiang:2024cpe, Yang:2024cnd, Meng:2024cnq, Li:2024tld, QiQi:2024dwc}. However, such dynamics are not isolated from the galactic environment: dark matter (DM) halos, which dominate the mass distribution in galaxies \cite{Navarro:1995iw,Hernquist:1990be,Gondolo:1999ef}, can imprint measurable modifications on the SMBH's spacetime \cite{Cardoso:2021wlq,Sadeghian:2013laa, Barausse:2014tra}.

DM halos, integral components of galaxies, significantly influence the dynamics of objects within them. These halos are often modeled using various density profiles, each designed to capture the diverse characteristics of galaxies. A generalized parametrization, frequently employed, expresses the density $\rho(r)$ as a function of radial distance ($r$) from the halo's center \cite{Taylor:2002zd}
\begin{equation}
\rho(r) = 2^{(\gamma - \alpha)/k} \rho_0 (r/a_0)^{-\alpha} (1 + (r/a_0)^k)^{-(\gamma - \alpha)/k},
\end{equation}
where $a_0$ represents the length-scale of the galaxy, $\rho_0$ is the DM density at $r=a_0$, and $\alpha$, $\gamma$, and $k$ are parameters defining the specific profile. These parameters reflect variations in the galaxy's size, mass, and shape. The profile plays a crucial role in shaping the gravitational potential near SMBHs \cite{Baes:2004pg}.

Several well-established models utilize this generalized form. For example, the Navarro-Frenk-White (NFW) profile \cite{Navarro:1995iw} commonly used for galaxies with significant DM content, sets $\alpha = 1$, $\gamma = 3$, and $k = 1$. The NFW profile describes a cuspy density distribution near the center, gradually declining at larger radii. For dwarf galaxies, the Burkert profile, which corresponds to $\alpha = 1$, $\gamma = 3$, and $k = 2$, is often more suitable \cite{Burkert:1995yz}. This profile features a core-like structure in the central region, avoiding the central cusp of the NFW profile. Another common model is the Hernquist profile, which corresponds to $\alpha = 1$, $\gamma = 4$, and $k = 1$, often used to describe the S$e$rsic profiles observed in elliptical galaxies and galactic bulges \cite{Hernquist:1990be}.

It is natural to explore that how the DM distributions can affect the spacetime of the central black hole. One approach involves modeling an isolated black hole spacetime matched to a specific matter distribution via the mass function \cite{Hou:2018bar, Xu:2018wow, Konoplya:2019sns, Jusufi:2020cpn, Liu:2021xfb, Zhang:2022roh, Zhang:2021bdr, Xu:2021dkv}. Another method employs relativistic extensions of Einstein clusters \cite{Einstein_1939, Cardoso:2021wlq, Ciotti:1996pf}, which provide a framework for constructing self-consistent spacetimes of SMBHs embedded in DM halos. In such models, the anisotropic pressure of the halo modifies the effective potential, affecting key orbital properties such as the innermost stable circular orbits (ISCOs) and light rings. For instance, Cardoso et al. \cite{Cardoso:2021wlq} demonstrated that DM halos can induce gravitational redshift and alter GW propagation, resulting in detectable phase shifts in EMRIs' signals. In their study, Cardoso et al. \cite{Cardoso:2021wlq} utilized a generalized version of the ``Einstein cluster" to develop an exact field solution for a black hole immersed in DM. Their model adopted a Hernquist-type density distribution for the DM halo, given by:
\begin{equation}\label{1}
\rho(r) = \frac{M_H a_0}{2 \pi r (r + a_0)^3}.
\end{equation}
Using similar methods, spacetime solutions for black holes embedded in DM with alternative density profiles have been constructed in a series of works \cite{Konoplya:2022hbl, Jusufi:2022jxu, Shen:2023erj, Shen:2024qxv, Maeda:2024tsg, Al-Badawi:2024asn}, extending the applicability of such models to a broader range of galactic environments. Numerical approach to black hole solutions with generic dark matter profiles and their implications in GW radiations have also been explored in ref.~\cite{figueiredo2023black}.

While periodic orbits in vacuum black hole spacetimes are well-studied, the role of DM halos remains unexplored. In this paper, we analyze periodic orbits and their GW signatures for SMBHs surrounded by Hernquist-like DM halos \cite{Hernquist:1990be}. Using the balck hole solution sorrounding by the DM halo derived in \cite{Cardoso:2021wlq} from a generalized Einstein cluster model \cite{Einstein_1939}, we derive the effective potential for timelike geodesics and compute corrections to marginally bound orbits (MBOs) and ISCOs. We show that the halo's length scale $a_0$ and central density $\rho_0$ reduce orbital radii and angular momenta, shifting the $(z,w,v)$ taxonomy of periodic orbits. Employing the numerical ``kludge" framework \cite{Babak:2006uv}, we simulate GW waveforms for representative orbits, revealing DM-induced phase shifts and amplitude modulations. We then conduct a detailed analysis of the frequency spectra of the gravitational waveforms arising from these periodic orbits and evaluated their potential detectability by comparing their characteristic strains with the sensitivity curve of space-based GW detector, LISA.

The structure of this paper is as follows: In Sect. \ref{spacetime}, we provide a brief overview of the black hole spacetime model surrounded by Hernquist-like DM halos. Sect. \ref{orbits} focuses on the geodesics of a small-mass object orbiting around SMBHs within such DM halos. The potential and bounded orbits of these objects are also discussed. While Sect. \ref{periodic} offers a detailed classification of periodic orbits. Sect. \ref{GW} examines the GW waveforms associated with these dynamics. Finally, the results are summarized in Sect. \ref{conclusion}.

\section{Black hole spacetime in DM halo}\label{spacetime}
\renewcommand{\theequation}{2.\arabic{equation}} \setcounter{equation}{0}

This section presents a brief introduction to a spherically symmetric black hole in the Hernquist-like DM halo. The line element of this black hole is given by \cite{Cardoso:2021wlq, figueiredo2023black}
\begin{equation}\label{2}
ds^2= - a(r)dt^2+\frac{dr^2}{b(r)}+r^2d\theta^2+r^2 \sin^2\theta  d\phi^2,
\end{equation}
where 
\begin{eqnarray}\label{4}
a(r)&=&\bigg(1- 2 \frac{M_{B}}{r}\bigg)e^{\gamma},\\
b(r)&=&1- 2 m(r)/r,
\end{eqnarray}
with
\begin{eqnarray}\label{5}
\gamma&=&-\pi \sqrt{\frac{M_H}{\xi}}+2 \sqrt{\frac{M_H}{\xi}}\tan^{-1}\bigg(\frac{r+a_0-M_H}{\sqrt{M_H\xi}}\bigg),\label{6} \nonumber\\
&& \\
\xi&=&2 a_0 - M_H + 4 M_{B}.
\end{eqnarray}
The radial function $a(r)$ acquires asymptotic flatness at large distances and is related to the mass function $m(r)$ by the relation
\begin{equation}
r \frac{a^\prime(r)}{a(r)}=\frac{m(r)}{r-2 m(r)}.
\end{equation}
Since we are working in Hernquist-type profile, the $m(r)$ is chosen to be 
\begin{equation}\label{3}
m(r)=M_{B}+ \frac{M_H r^2}{(a_0+r)^2}\bigg(1- \frac{2 M_{B}}{r}\bigg)^2.
\end{equation}
which, at small distances, describes a black hole with mass $M_{B}$ while at large scales it corresponds to a distribution of matter described by Eq.~(\ref{1}). This spacetime, similar to the Schwarzschild black hole, possesses an event horizon at $r = 2M_{BH}$ and a curvature singularity at $r = 0$. This profile corresponds to matter density 
\begin{equation}
\rho_{\rm DM}(r)=\frac{m^{\prime}}{4\pi r^2}=\frac{2 M_H (a_0+2 M_{B})(1-2 M_{B}/r)}{4\pi r(r+a_0)^3}.
\end{equation}
At large distances, specifically when $a_0\gg r \gg M_B$, the DM density can be approximated as
\begin{equation}
\rho_{\rm DM} \sim \frac{M_H}{a_0^2}\frac{1}{r}, 
\end{equation}
which indicates that for a fixed value of $M_H$ and large $a_0$ or for increasing $r$ the DM density $\rho_{\rm DM}$ decreases.
The ADM mass of this black hole here is the sum of the mass of the DM halo and that of the black hole, i.e. $M_{ADM}= M_H + M_{B}$. It is also worth mentioning here that the matter density vanishes at the horizon and the tangential pressure is regular there. The ratio $M/a_0$ is termed the compactness parameter and can assume values greater than $10^{-4}$, in a galactic context \cite{Cardoso:2021wlq}.

\section{Geodesics, Effective potential, and bounded orbits}\label{orbits}
\renewcommand{\theequation}{3.\arabic{equation}} \setcounter{equation}{0}

We now examine the geodesic motion, effective potential, and bound orbits of a massive test particle within the gravitational field of a spherical black hole with a Hernquist-type DM halo given in the above section. The presence of the halo will affect the particle's trajectory.

\subsection{Geodesics}

The particle's motion is governed by the Lagrangian formalism, where the Lagrangian and generalized momentum, respectively, is given by
\begin{eqnarray}
\mathcal{L}&=&\frac{1}{2}g^{\mu\nu}\dot{x}^\mu\dot{x}^\nu=- m. \label{Lagrange}
\end{eqnarray}
The dot here means derivative with respect to the proper time $\tau$ and $m$ is the mass of the test particle. For simplicity, we set the particle's mass $m$ to unity. Then the generalized momentum $p_\mu$ of the particle can be obtained via
\begin{eqnarray}
p_\mu &=&\frac{\partial \mathcal{L}}{\partial\dot{x}^\mu}.\label{momenta}
\end{eqnarray}
From Eqs.~(\ref{Lagrange}) and (\ref{momenta}), one finds
\begin{eqnarray}
p_t &=& -a(r) \dot{t}= -E,\\
p_r&=&\frac{\dot{r}}{b(r)},\\
p_\theta &=&r^2 \dot{\theta},\\
p_\phi &= &r^2 \sin^2\theta \dot{\phi}=L,
\end{eqnarray}
where two conserved quantities govern the particle's motion: energy $E$ and orbital angular momentum $L$ of the particle's per unit mass. A massive particle moving in the spherically symmetric background of a black hole can be constrained to the equatorial plane, $\theta = \pi/2$ and $\dot{\theta}=0$. We then can compute the geodesics using Eqs. (\ref{Lagrange}) and (\ref{momenta}), one obtains
\begin{eqnarray}
\dot{t}&=&
\frac{E}{a(r)},\\
\dot{\phi}&=&
\frac{L}{r^2},
\end{eqnarray}
and
\begin{equation}\label{r2}
\dot{r}^2={b(r)}\bigg(-1+\frac{E^2}{a(r)}-\frac{{L^2}}{r^2}\bigg).
\end{equation}

\subsection{Effective potential and bounded orbtis}

The primary focus of this study lies in examining the characteristics of periodic orbits around a black hole with a DM halo. These orbits are a subset of bound orbits confined within a specific range of energy and angular momentum. The bounded orbits can be stable or unstable. The effective potential is a key tool for understanding orbital stability. Stable orbits reside at the potential's minima, while unstable orbits correspond to maxima.

\begin{figure*}
\centering
\includegraphics[width=8.8cm]{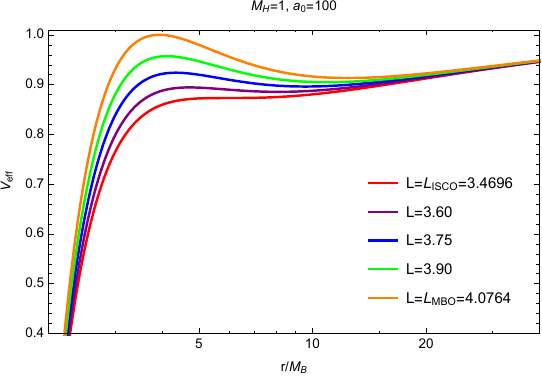} 
\includegraphics[width=8.8cm]{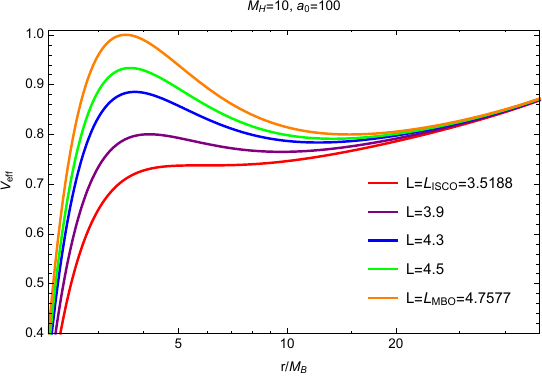} 
\includegraphics[width=8.8cm]{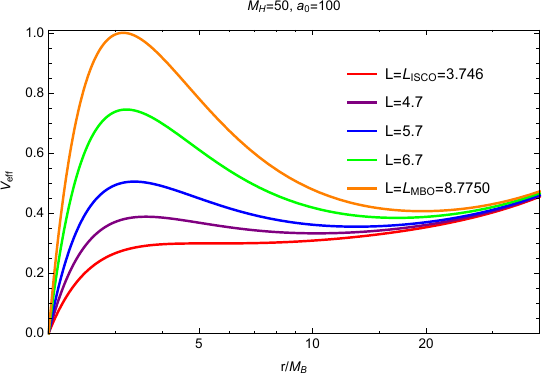} 
\includegraphics[width=8.8cm]{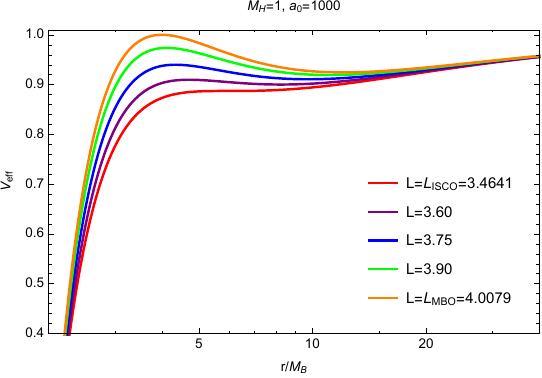} 
\includegraphics[width=8.8cm]{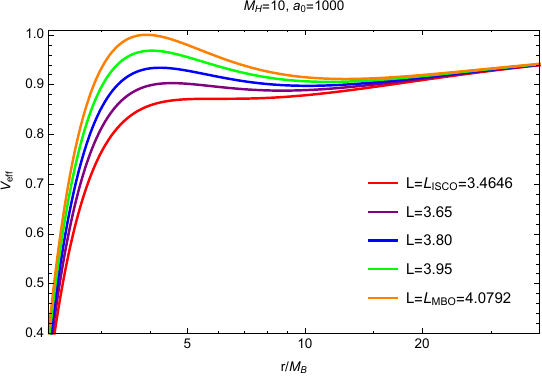}
\includegraphics[width=8.8cm]{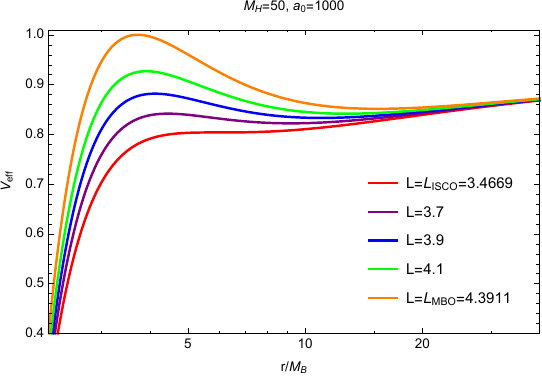}
\caption{The effective potential of the test particle moving around a black hole with DM halo.}
\label{effective}
\end{figure*}

To study effective potential, the radial Eq. (\ref{r2}) can be written as
\begin{equation}
\dot{r}^2=\frac{b(r)}{a(r)}\bigg(E^2-V_{\rm eff} \bigg),
\end{equation}
where
\begin{equation}\label{ep}
V_{\rm eff}=a(r)\left(1+\frac{L^2}{r^2}\right)
\end{equation}
is the radial effective potential which describes the radial motion of the particle. Increasing angular momentum changes the effective potential, creating extrema. This behavior is illustrated in Fig.~\ref{effective} where it is observed that as the angular momentum increases, the value of $V_{\rm eff}$ also increases. For fixed $a_0$, as  extrema of $V_{\rm eff}$ decreases with increasing $M_H$. Conversely, for a fixed $M_H$, the extrema of $V_{\rm eff}$ increase with increasing $a_0$. Since the function $a(r)$ admits asymptotic flatness, thus $V_{\rm eff}\rightarrow 1$ as $r\rightarrow +\infty$. So $E=1$ is the critical energy of the particle moving in an orbit. The particle having energy $E>1$ will move with positive velocity at infinity i.e. $\dot{r}^2>0$ as $E^2-{V_{\rm eff}}_{r\rightarrow +\infty}>0$, implying the existence of unbounded orbits. Therefore the maximum energy the particle can acquire while moving in a bounded orbit must be $E\le 1$.

For a massive particle around a spherically symmetric black hole immersed in DM halo, bounded orbits exist between MBOs and ISCOs. Between these two points, there exists a range of energies that allow for bound orbits to exist. 
The energy of a particle in a bound orbit around a black hole must indeed lie within the range $E_{\rm ISCO} \le E \le E_{\rm MBO} = 1$. $E_{\rm MBO}=1$ corresponds to the energy of the MBO, where the particle has just enough energy to escape to infinity. $E_{\rm ISCO}$ represents the energy of the ISCO, and any particle with energy less than this will plunge into the black hole. The condition for angular momentum is slightly different. The angular momentum $L$ of a particle in a bound orbit must satisfy $L \ge L_{\rm ISCO}$, where $L_{\rm ISCO}$ being the angular momentum of the particle in ISCO.

\subsection{Marginally stable orbit}

\begin{figure}
\centering
\includegraphics[width=8.8cm]{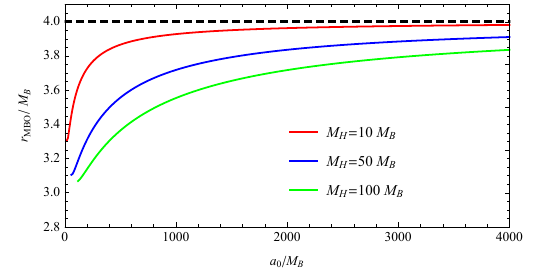} 
\includegraphics[width=8.8cm]{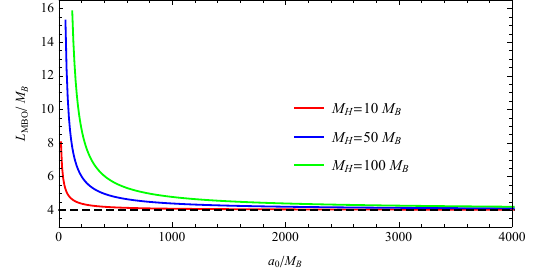}
\captionsetup{justification=raggedright,singlelinecheck=false}
\caption{The radius $r_{\rm MBO}$ (upper panel) and angular momentum $L_{\rm MBO}$ (lower panel) for MBOs of a test particle around a black hole with DM halo with respect to parameter $a_0$, for different values of $M_H$.}
\label{rmbo}
\end{figure}

The MBO represents the smallest possible circular bound orbit, possessing the minimum radius and an energy of $E_{\rm MBO} = 1$. Given the effective potential $V_{\rm eff}$ in Eq.~(\ref{ep}), governing the particle's motion, the MBO satisfies the conditions 
\begin{enumerate}[(i)]
\item $V_{\rm eff} = 1$  
\item $dV_{\rm eff}/dr = 0$.
\end{enumerate}
Solving these equations yields the radius and angular momentum of the MBO. 
While exact solutions cannot be computed due to the complexity of the equations, thus approximated solutions are calculated by considering $M/a_0$ as a small quantity, we get the radius for the MBO $r_{\rm MBO}$ as
\begin{eqnarray}
r_{\rm MBO}\simeq \tilde{r}_{\rm MBO}\left[1- 2 \frac{M_H}{a_0}+\frac{4 M_H}{3 a_0^2}\big(13 M_H + 12 M_{B})\right], \nonumber\\
\end{eqnarray}
and the corresponding angular momentum, $L_{\rm MBO}$, as
\begin{eqnarray}
L_{\rm MBO}\simeq \tilde{L}_{\rm MBO}\left[1+ 2 \frac{M_H}{a_0}-\frac{4 M_H}{3 a_0^2}\big(M_H+6 M_{B}\big)\right]. \nonumber\\
\end{eqnarray}
For a Schwarzschild black hole, the MBO radius and angular momentum are given by $\tilde{r}_{\rm MBO} = 4 M_B$ and $\tilde{L}_{\rm MBO} = 4 M_B$, respectively, where $M_B$ denotes the black hole mass. The radius $r_{\rm MBO}$ and the angular momentum $L_{\rm MBO}$ of the MBO exhibit distinct dependencies on the DM halo parameters $a_0$ and $M_H$.  Fig.~\ref{rmbo} illustrates the numerical results of $r_{\rm MBO}$ and $L_{\rm MBO}$ as functions of $a_0$ and $M_H$.  We observe that $r_{\rm MBO}$ increases with increasing $a_0$, while $L_{\rm MBO}$ exhibits the opposite trend, decreasing as $a_0$ increases. Conversely, as $M_H$ increases, $r_{\rm MBO}$ decreases and $L_{\rm MBO}$ increases.

\subsection{Innermost stable circular orbits}

\begin{figure}
\centering
\includegraphics[width=8.8cm]{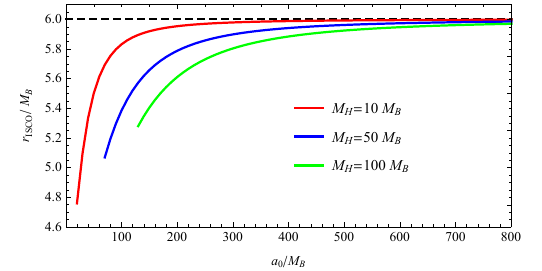} 
\includegraphics[width=8.8cm]{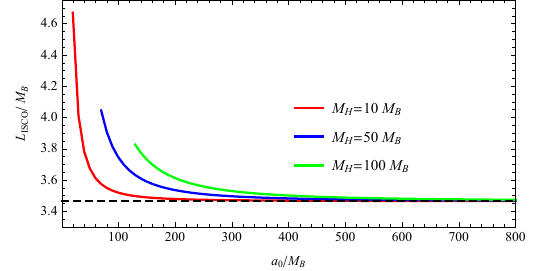}
\includegraphics[width=8.8cm]{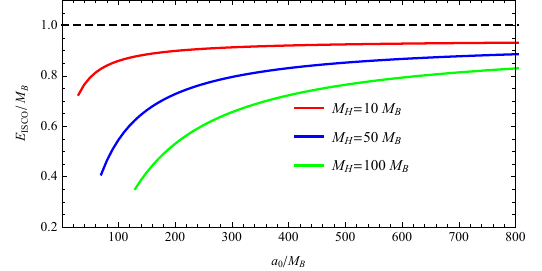}
\captionsetup{justification=raggedright,singlelinecheck=false}
\caption{The radius $r_{\rm ISCO}$ (upper panel), the angular momentum $L_{\rm ISCO}$ (middle panel), and the energy $E_{\rm ISCO}$ (lower panel) for the ISCOs. }
\label{isco}
\end{figure}

\begin{figure}
\centering
\includegraphics[width=8.8cm]{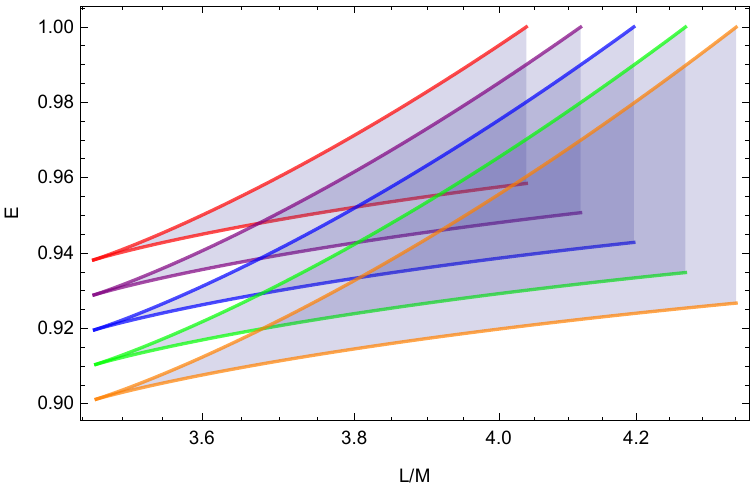} 
\includegraphics[width=8.8cm]{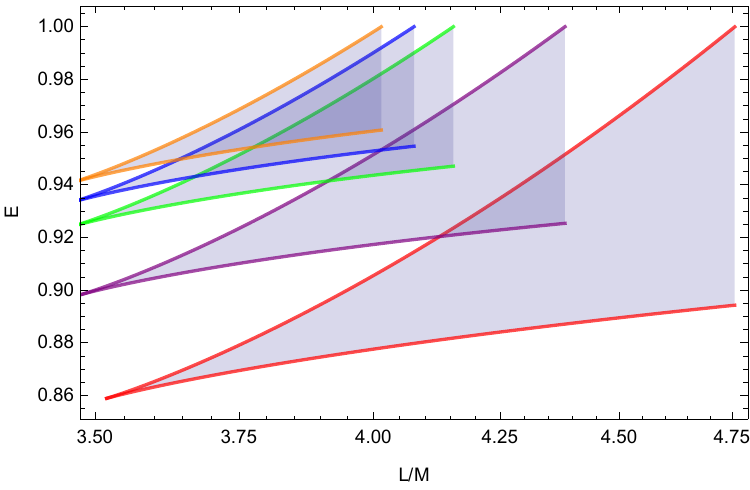}
\captionsetup{justification=raggedright,singlelinecheck=false}
\caption{The allowed $L-E$ parametric space for a timelike particle orbiting around a black hole with DM halo. \textbf {Upper panel}: The value of $a_0$ is kept fixed at $a_0= 1000 M_B$ while ${M_H}/{M_B}$= 5,15, 25, 35, 45 (\textit{top to bottom}). \textbf{Lower panel}: The value of $M_H=10 M_B$ is kept fixed while $a_0/M_B$= 100, 200, 500, 1000, 5000 (\textit{bottom to top}).}
\label{ELspace}
\end{figure}

As the name indicates, the ISCOs are the ones beyond which the particle will plunge into the black hole. These orbits fulfill the following conditions.
\begin{enumerate}[(i)]
	\item $V_{\rm eff}=E^2$,
	\item $\partial_r V_{\rm eff}=0$,
	\item $ \partial_{rr} V_{\rm eff}=0$.
\end{enumerate}
By solving these three conditions using the effective potential that includes the DM halo's contribution, we can determine the precise location of the ISCO in our model and quantify the halo's influence on the orbital dynamics. From (i) and (ii) we obtain the angular momentum of the particle in ISCO as
\begin{equation}\label{Lisco}
L_\text{ISCO}=\sqrt{\frac{r^3 a^\prime(r_{\rm ISCO})}{2 a(r_{\rm ISCO})-r a^\prime(r_{\rm ISCO})}},
\end{equation}
and energy as
\begin{equation}\label{Eisco}
E_\text{ISCO}=\sqrt{\frac{2 a^2(r_{\rm ISCO})}{2 a(r_{\rm ISCO}) - r_{\rm ISCO} a^\prime(r_{\rm ISCO})}}.
\end{equation}
From condition (iii) we get 
\begin{equation}
 a(r) a^{\prime\prime}(r)- 2  \big(a^\prime(r)\big)^2+\frac{3}{r} a(r) a^\prime(r)=0.
\end{equation}
Again, the presence of DM halo does not allow us to compute exact solutions to the above equations. Thus the approximated solutions are found by treating $M_H/a_0$ as small quantity,
\begin{eqnarray}
r_\text{ISCO}&\simeq &\tilde{r}_\text{ISCO}\left(1- 32\frac{M_H}{a_0^2}M_{B}\right), \\
L_\text{\tiny ISCO} &\simeq& \tilde{L}_\text{\tiny ISCO}\left(1+16 \frac{M_H}{a_0^2}M_{B}\right).
\end{eqnarray}
where $\tilde{r}_\text{\tiny ISCO}=6 M_B$ and $\tilde{L}_\text{\tiny ISCO}=2 \sqrt{3} M_B$ are the particle's radius and angular momentum at ISCO's around the Schwarzschild black hole. In Fig.~\ref{isco}, for a fixed value of $M_H$, as $a_0$ increases, both $r_{\rm ISCO}$ and $E_{\rm ISCO}$ increase while $L_{\rm ISCO}$ decreases. Conversely, when $a_0$ is fixed, increasing $M_H$ leads to a decrease in $r_{\rm ISCO}$ and $E_{\rm ISCO}$, while an increase in $L_{\rm ISCO}$. The allowed $E$-$L$ regions for the existence of bound orbits are plotted in Fig.~\ref{ELspace}. For fixed values of halo's mass $M_H$ and length-scale $a_0$ energy $E$ changes with increase in angular momentum $L$. The width of these $E$-$L$ regions expands with increasing $M_H$ and contracts with increasing $a_0$.

\section{Periodic Orbits}\label{periodic}
\renewcommand{\theequation}{4.\arabic{equation}} \setcounter{equation}{0}

Having explored the ISCOs and MBOs, we now focus on periodic orbits near a black hole surrounded by DM halos. These bounded orbits are distinguished by their repetitive trajectories. Periodicity arises when the frequency ratio between the radial ($\omega_r$) and azimuthal ($\omega_\phi$) oscillations is rational. While a generic orbit may exhibit an irrational frequency ratio $\frac{\omega_\phi}{\omega_r}$, a rational number can always approximate this ratio with arbitrary precision.

This ability to approximate generic orbits with nearby periodic counterparts makes the study of periodic orbits invaluable for understanding broader orbital dynamics and their associated GW emissions. Such analysis provides critical insights into the complex GW signals produced by these systems, offering a robust framework for their interpretation.

Since the geometry under consideration is spherically symmetric thus the motion is entirely governed by radial $r$-motion and angular $\phi$-motion. The apsidal angle, $\Delta\phi$ can be calculated by integrating the change in azimuthal angle, $\phi$, over one radial oscillation, from the minimum radial distance $r_1$ (periapsis) to the maximum radial distance $r_2$, (apoapsis) and back, as
\begin{eqnarray}
\Delta\phi=\oint\, \, d\phi &=&2\int_{\phi_1}^{\phi_2}\,d\phi \nonumber\\
& = & 2 \int_{r_1}^{r_2}\,\frac{\dot{\phi}}{\dot{r}}dr.\nonumber
\end{eqnarray}
Using the radial Eq.~(\ref{r2}), we get
\begin{equation}\label{apsidal}
\Delta\Phi= 2 \int_{r_1}^{r_2} \frac{L}{r^2} \sqrt{\frac{a(r)}{b(r)\big(E^2-V_{\rm eff}\big)}} \,dr.
\end{equation}
Here the factor of 2 arises due to the symmetrical nature of the path followed by the massive particle. The apsidal angle $\Delta\phi$ traversed by the particle is significantly affected not just by the energy and angular momentum of the particle but also the structure of the spacetime around the black hole as the presence of the functions $a(r)$ and $b(r)$ depicts. Consequently, black holes with different values of parameters will exhibit varying apsidal angles.

\begin{figure*}
\centering
\includegraphics[width=8.5cm]{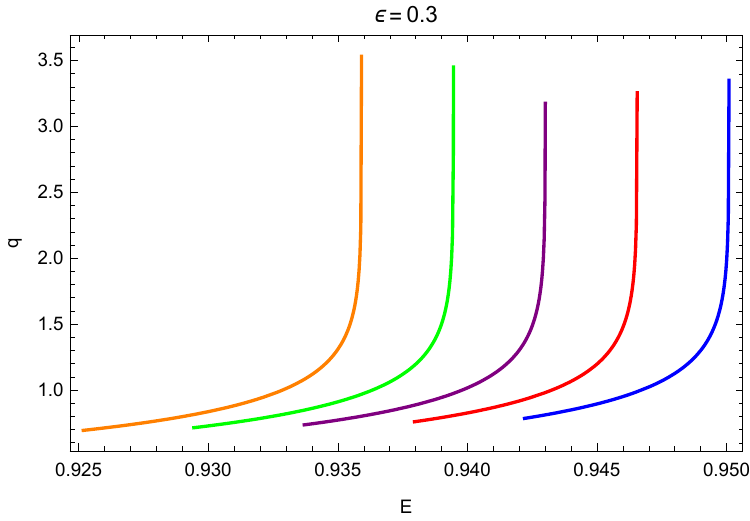} 
\includegraphics[width=8.5cm]{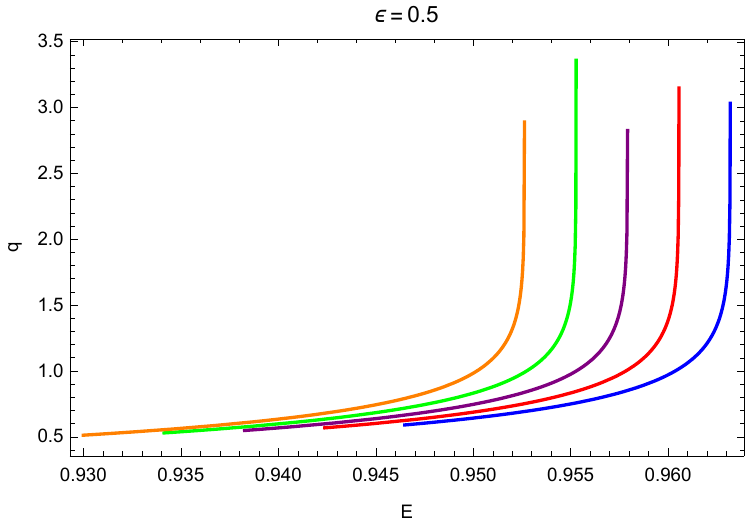}
\includegraphics[width=8.5cm]{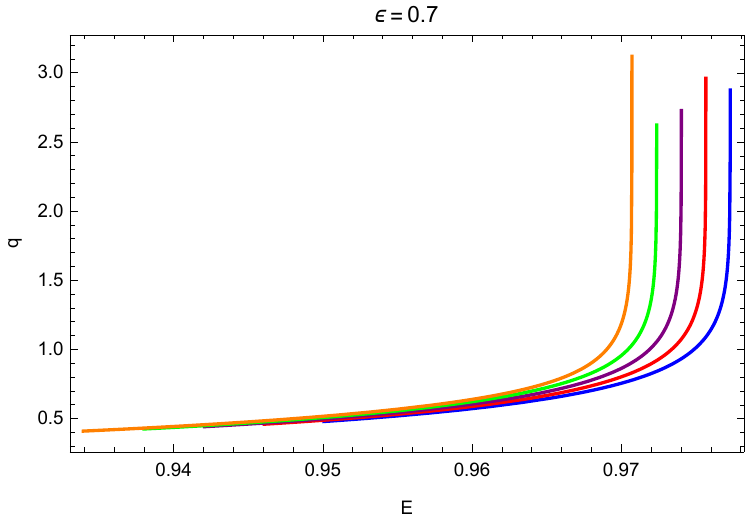} 
\includegraphics[width=8.5cm]{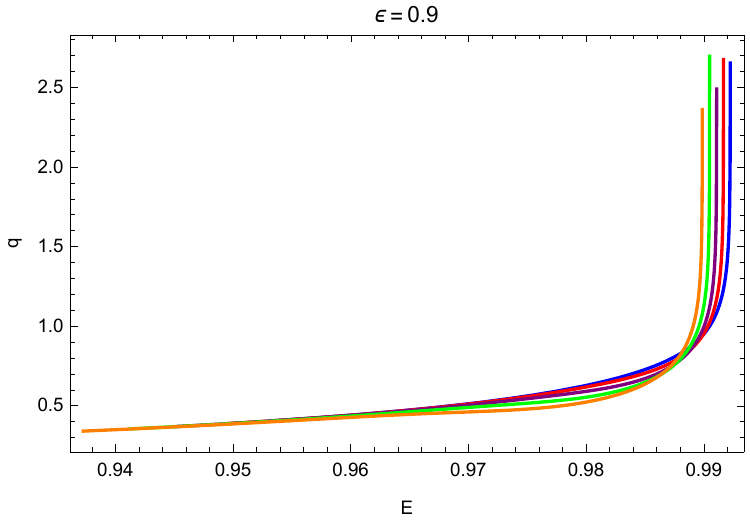} 
\caption{The rational number $q$ w.r.t energy $E$ while $M_H/M_B$= 10, 15, 20, 25, 30 (\textit{right to left}) and $a_0=1000 M_B$.}
\label{qe}
\end{figure*}

\begin{figure*}
\centering
\includegraphics[width=8.5cm]{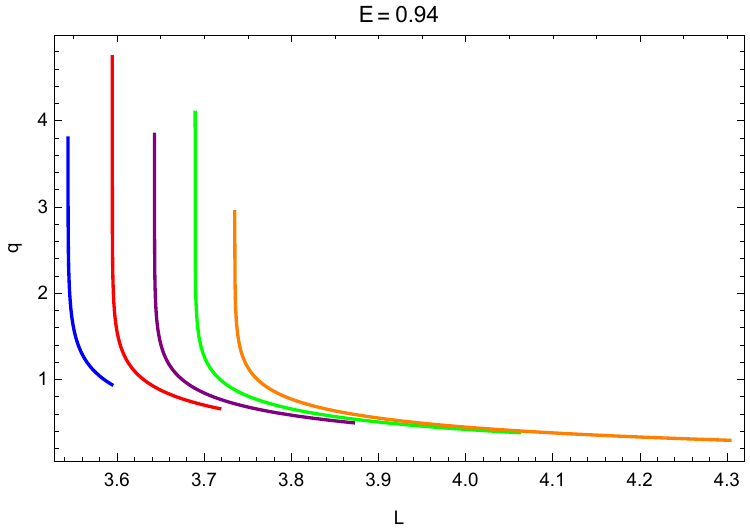}
\includegraphics[width=8.5cm]{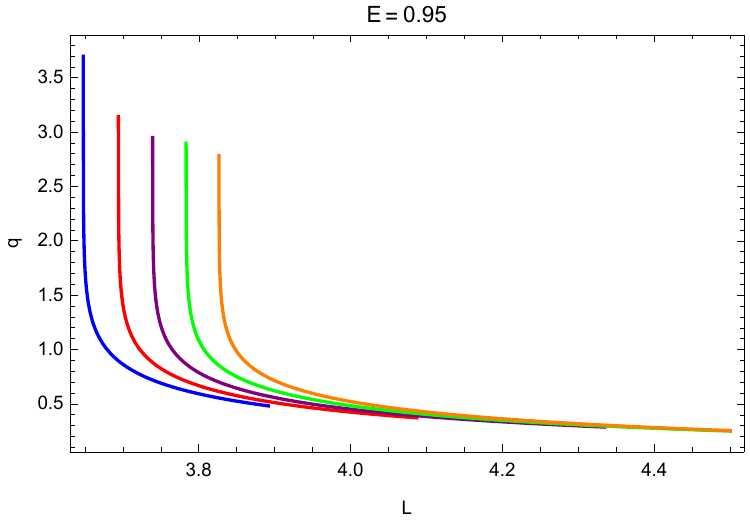} 
\includegraphics[width=8.5cm]{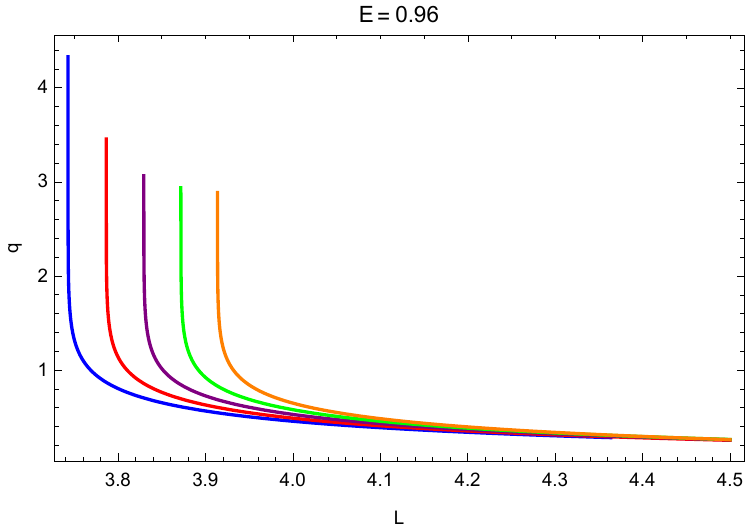}
\includegraphics[width=8.5cm]{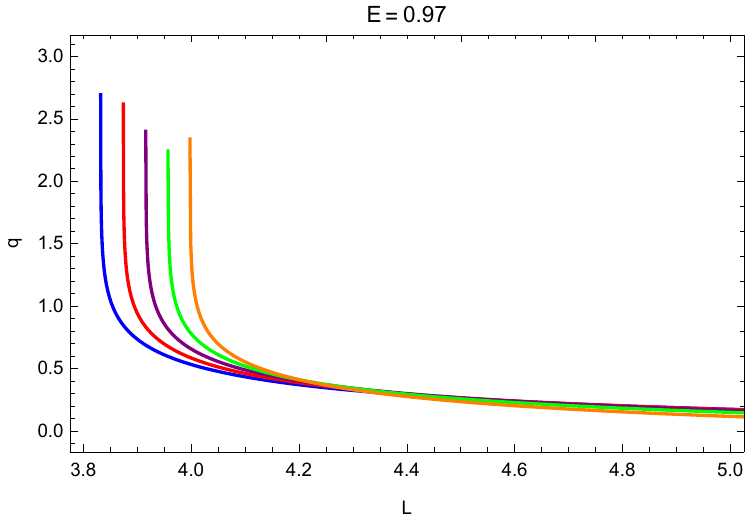} 
\captionsetup{justification=raggedright,singlelinecheck=false}
\caption{The rational number $q$ w.r.t orbital angular momentum $L$ while $M_H/M_B$= 10, 15, 20, 25, 30 (\textit{left to right}) and $a_0=1000 M_B$.}
\label{ql}
\end{figure*}

\begin{figure*}
\centering
\includegraphics[width=5.3cm]{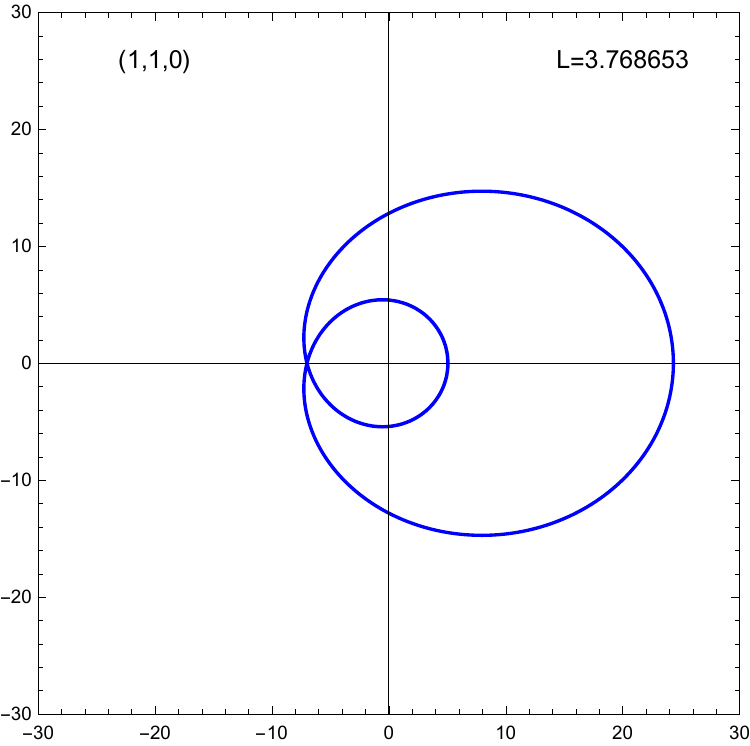} 
\includegraphics[width=5.3cm]{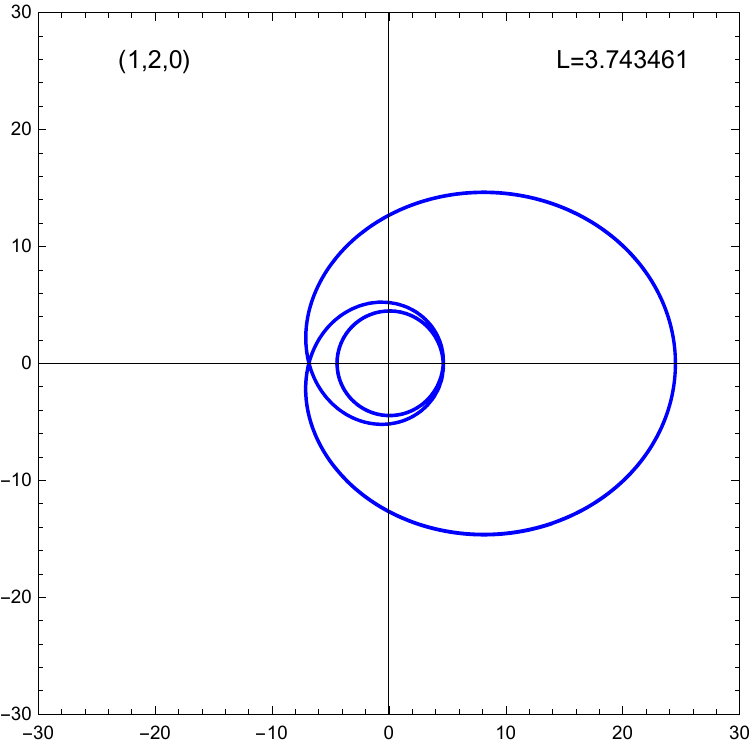}
\includegraphics[width=5.3cm]{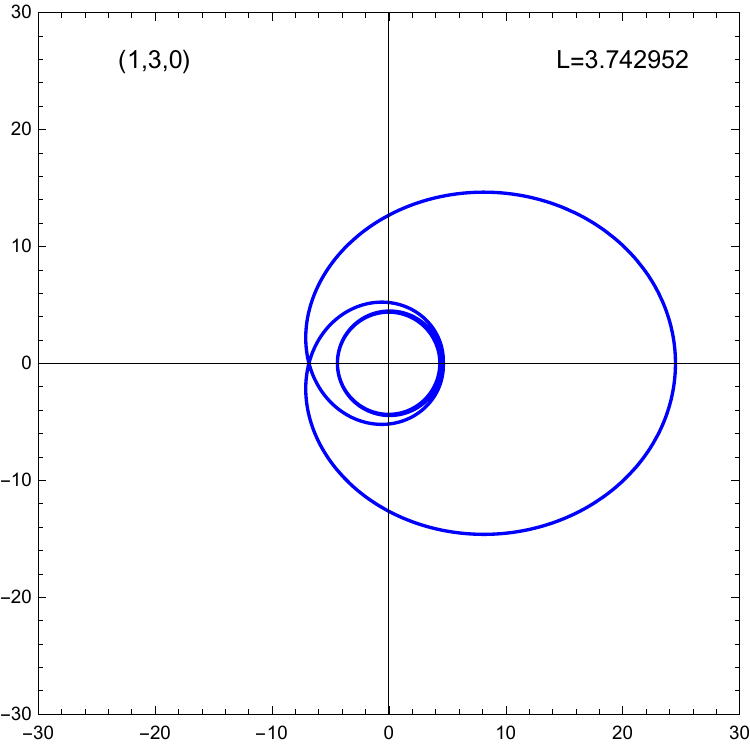} 
\includegraphics[width=5.3cm]{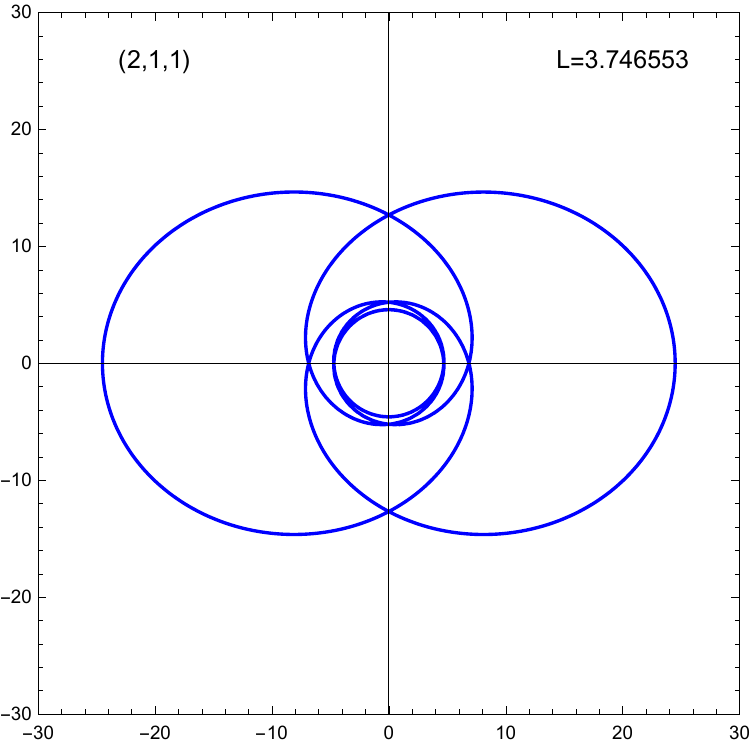} 
\includegraphics[width=5.3cm]{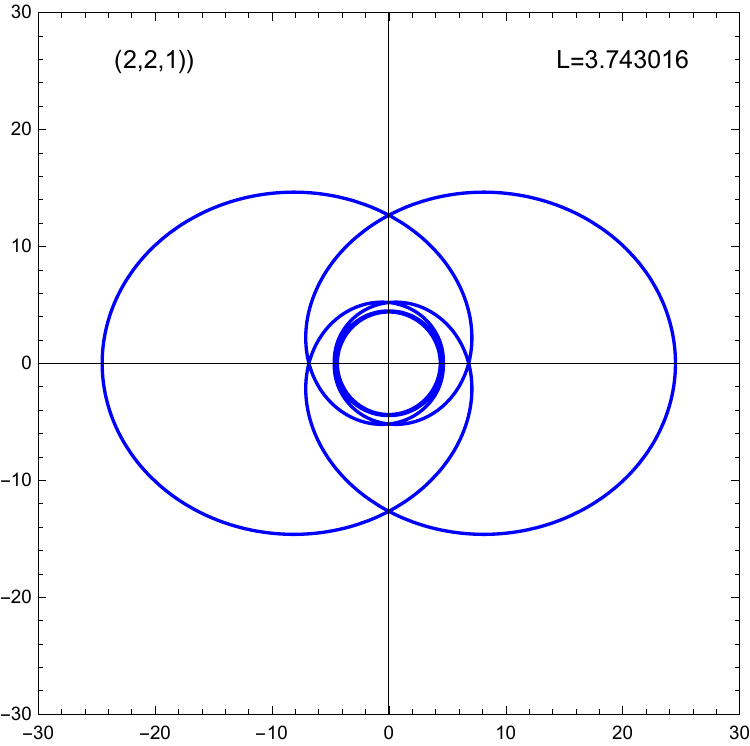} 
\includegraphics[width=5.3cm]{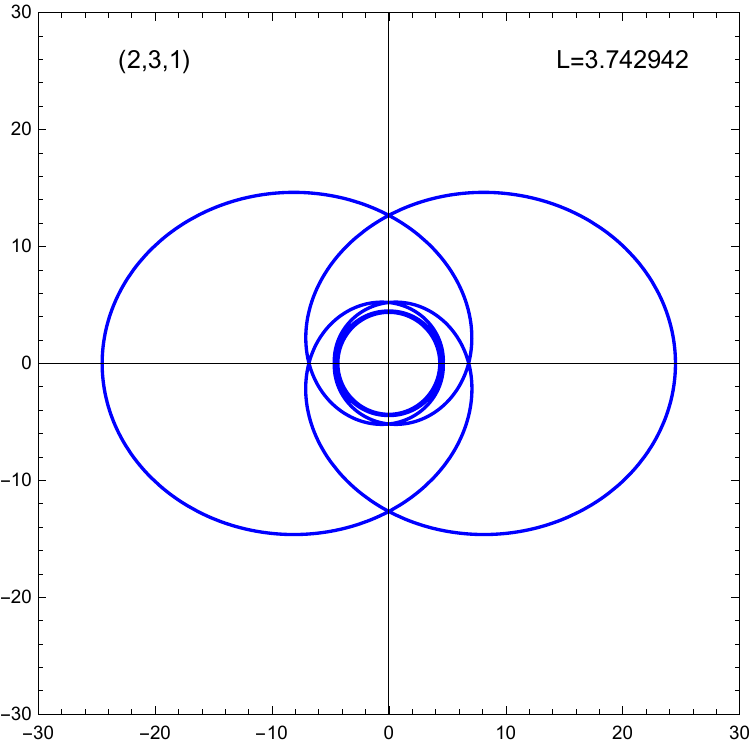} 
\includegraphics[width=5.3cm]{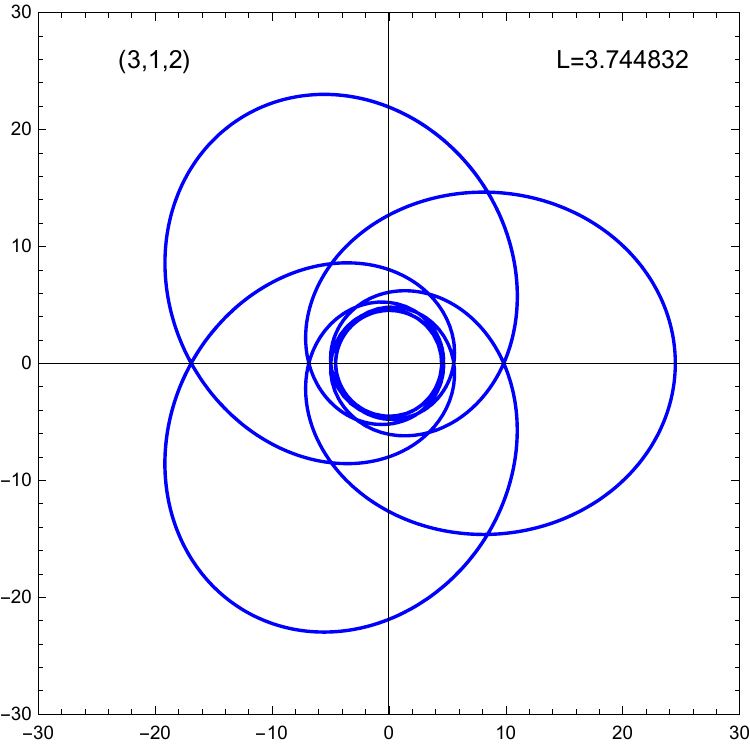} 
\includegraphics[width=5.3cm]{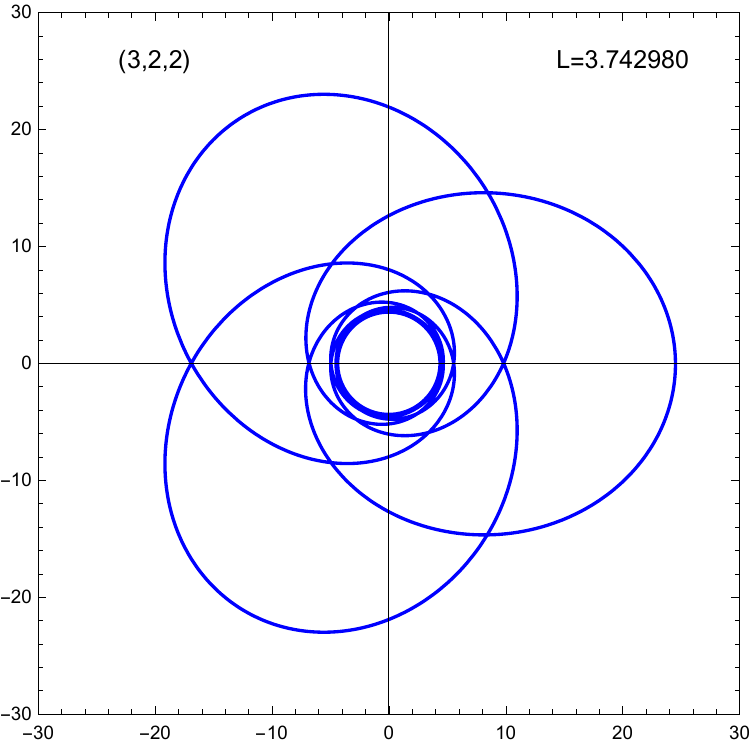}
\includegraphics[width=5.3cm]{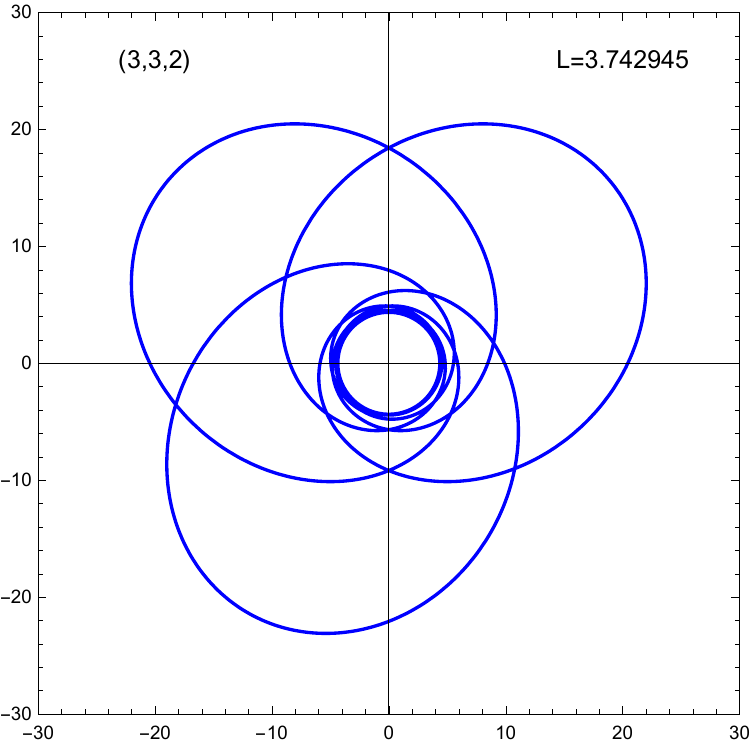} 
\includegraphics[width=5.3cm]{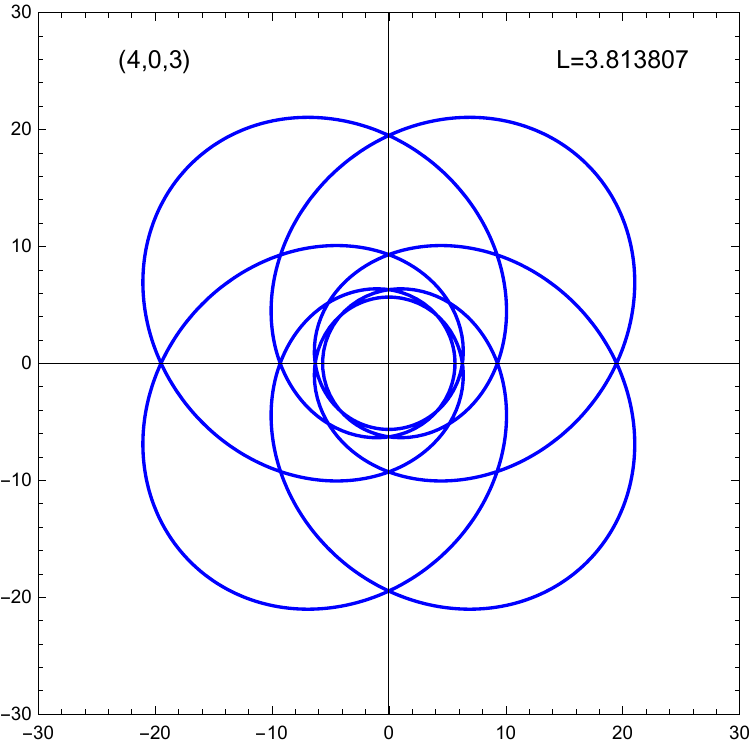} 
\includegraphics[width=5.3cm]{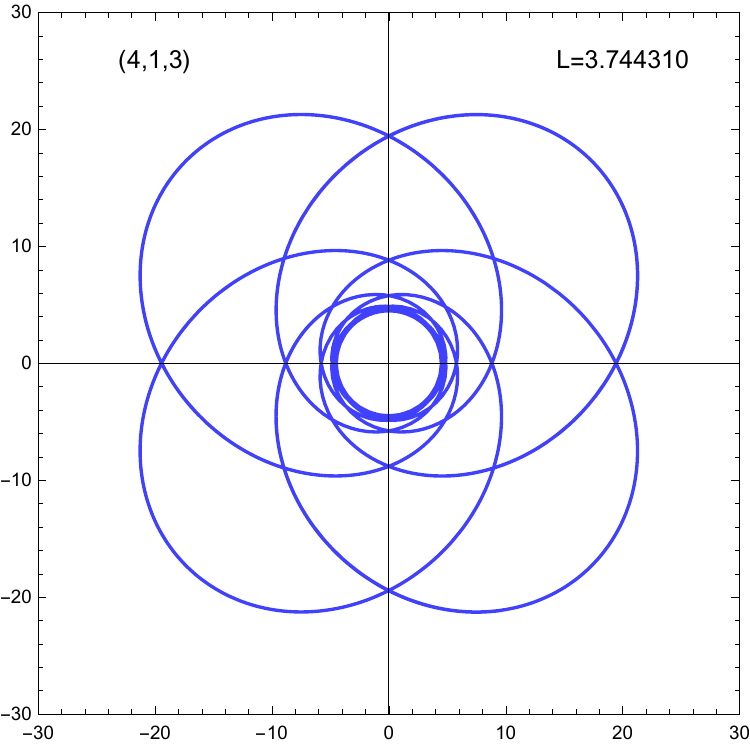} 
\includegraphics[width=5.3cm]{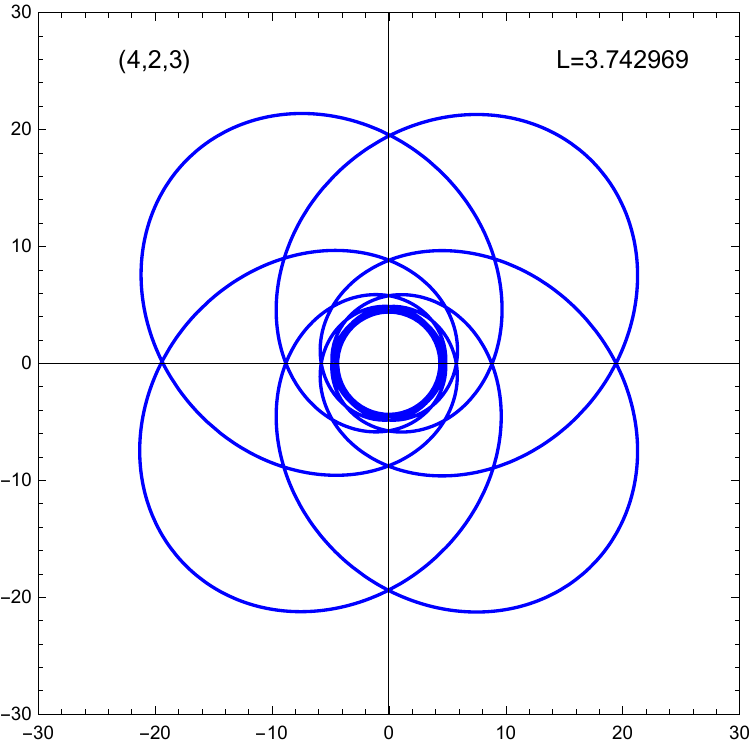} 
\captionsetup{justification=raggedright,singlelinecheck=false}
\caption{Periodic orbits for different values of $(z, w, v)$ around a black hole with DM halo. The value of parameters are $M=10 M_H$, $a_0=1000 M_H$ and $E=0.96$.}
\label{periodic1}
\end{figure*}

\begin{figure*}
\centering
\includegraphics[width=5.3cm]{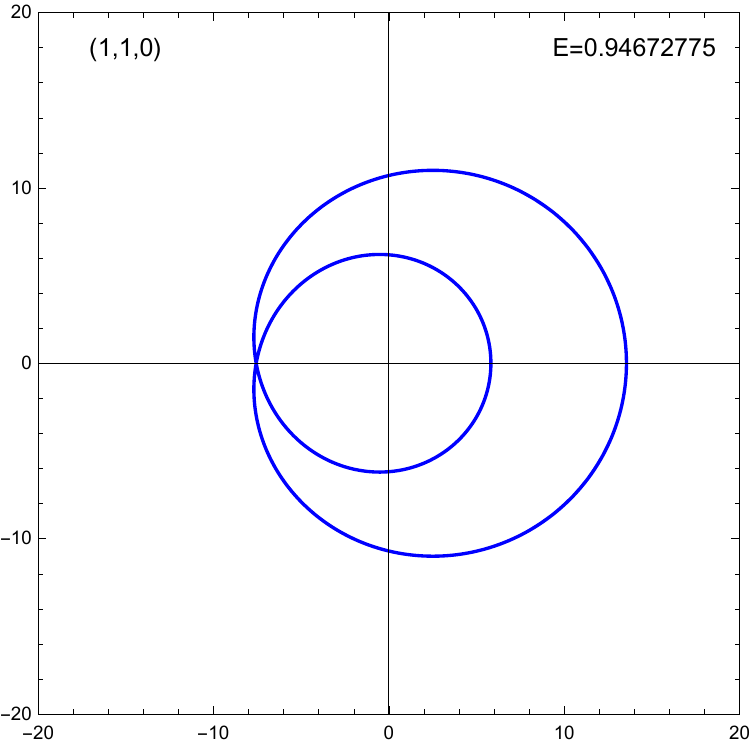} 
\includegraphics[width=5.3cm]{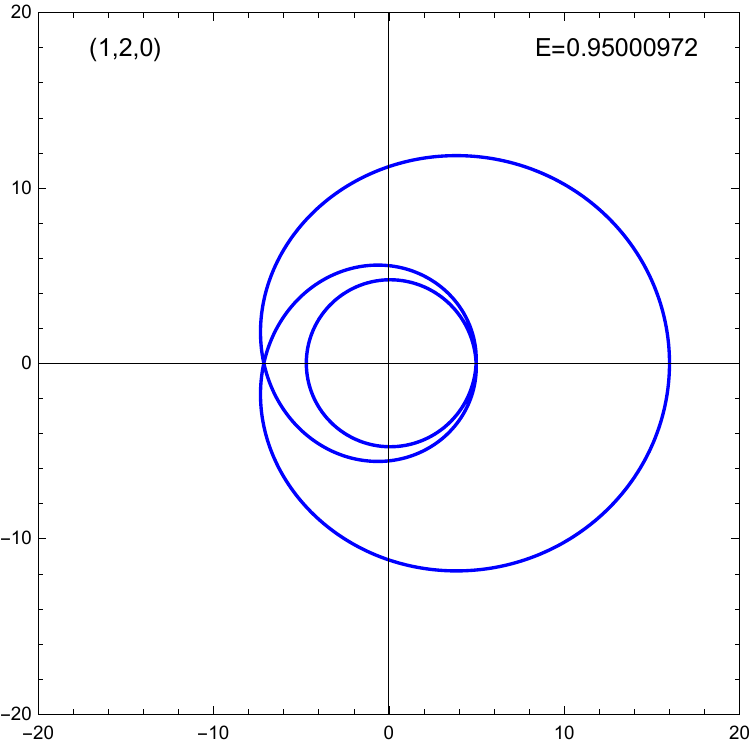}
\includegraphics[width=5.3cm]{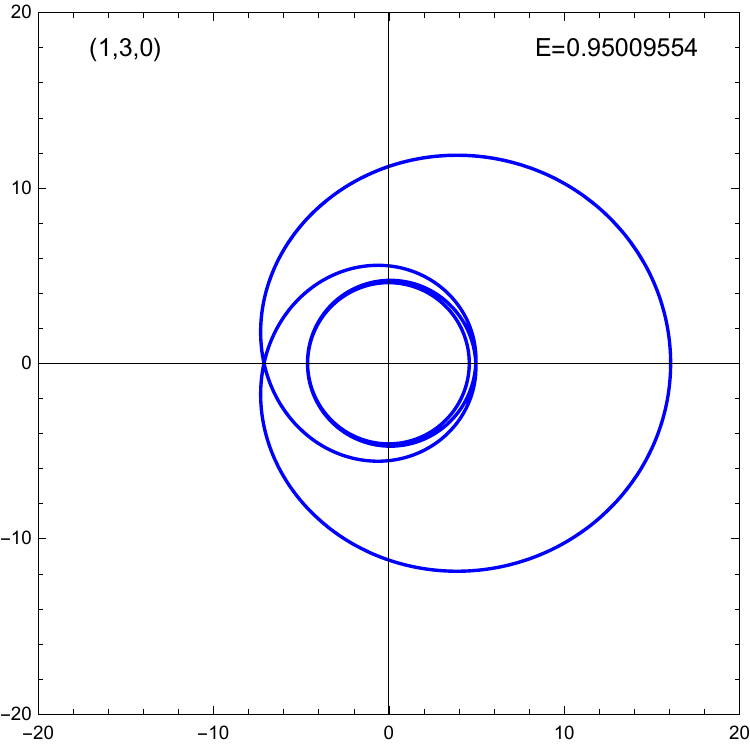} 
\includegraphics[width=5.3cm]{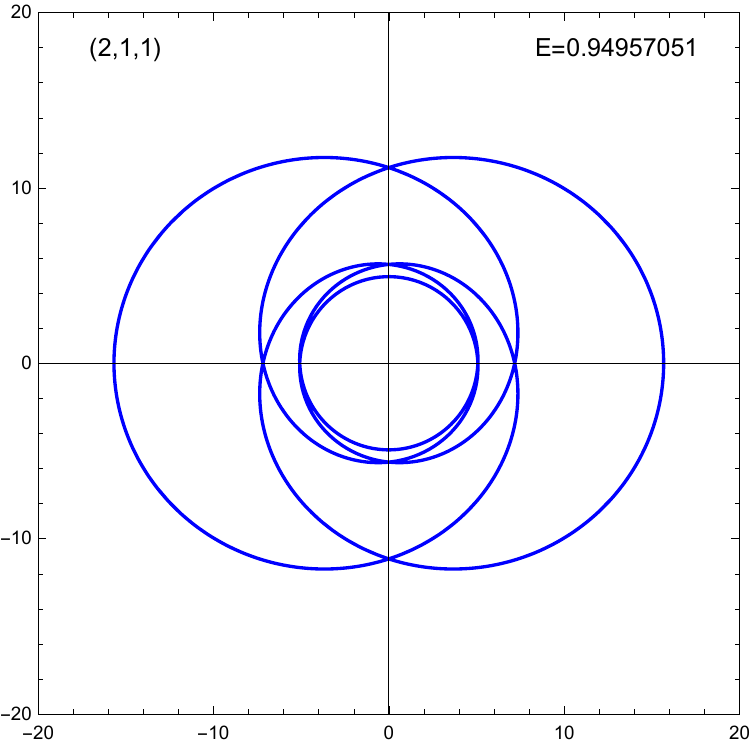} 
\includegraphics[width=5.3cm]{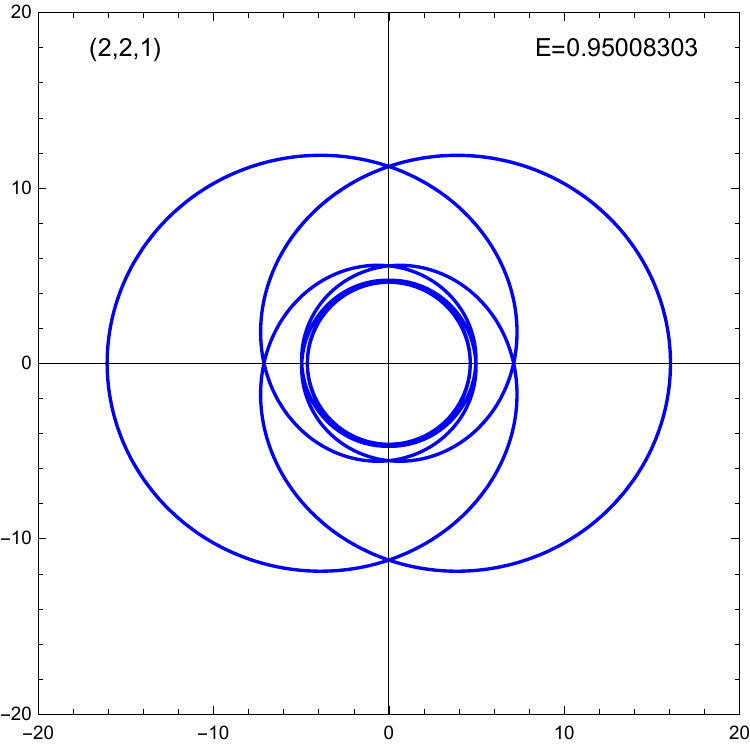} 
\includegraphics[width=5.3cm]{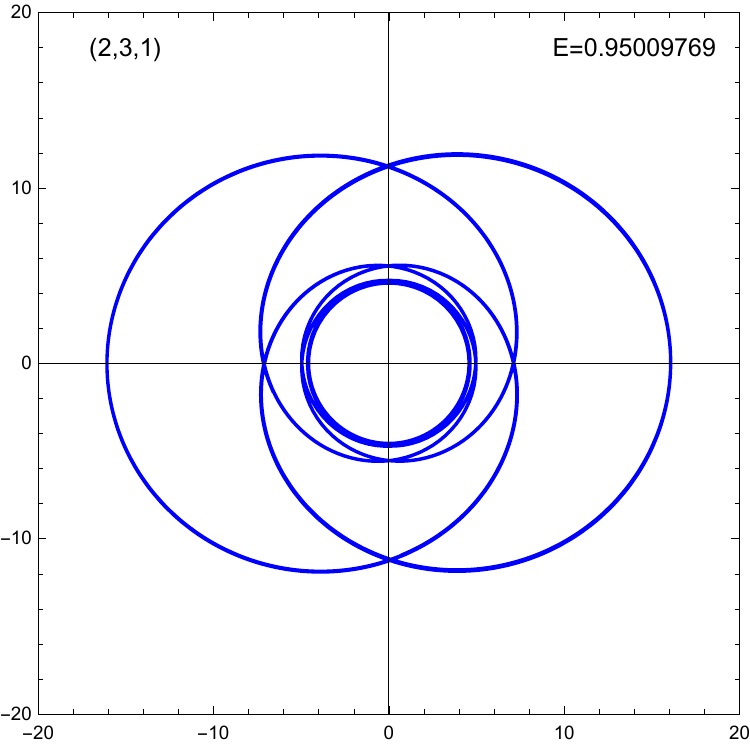} 
\includegraphics[width=5.3cm]{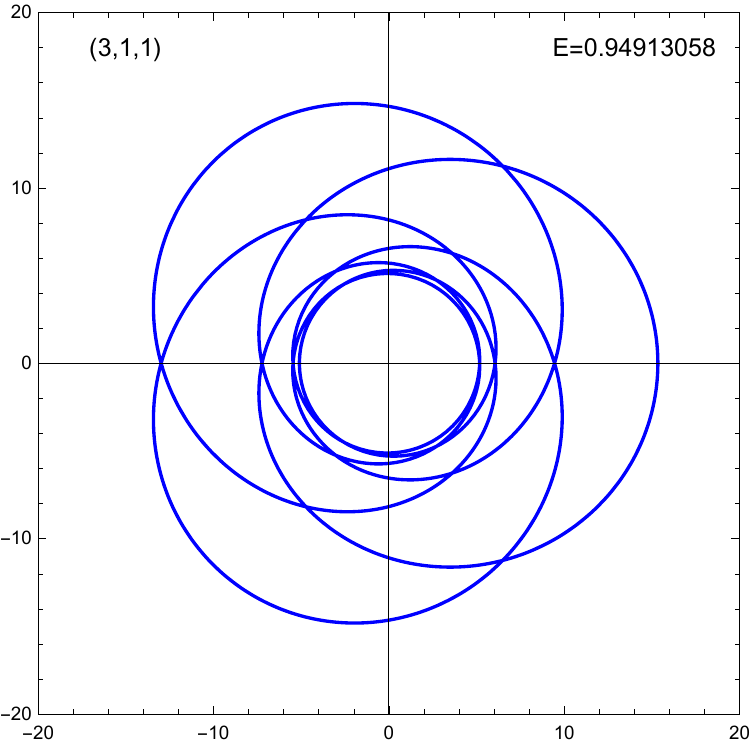} 
\includegraphics[width=5.3cm]{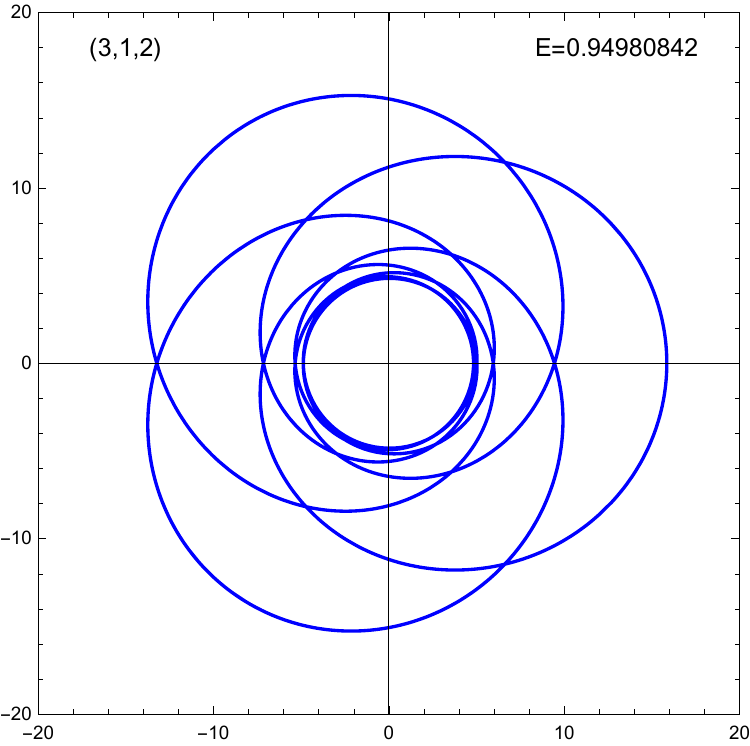} 
\includegraphics[width=5.3cm]{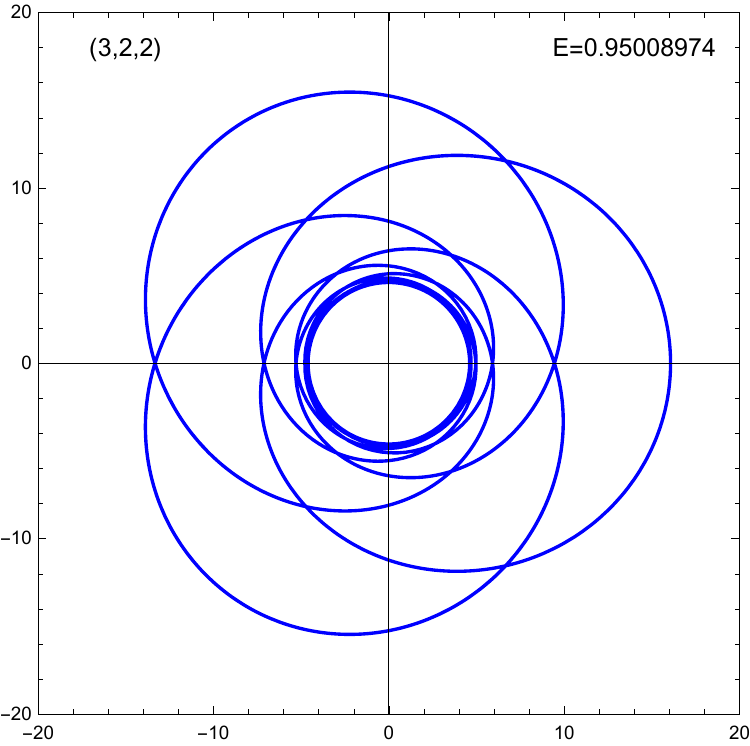}
\includegraphics[width=5.3cm]{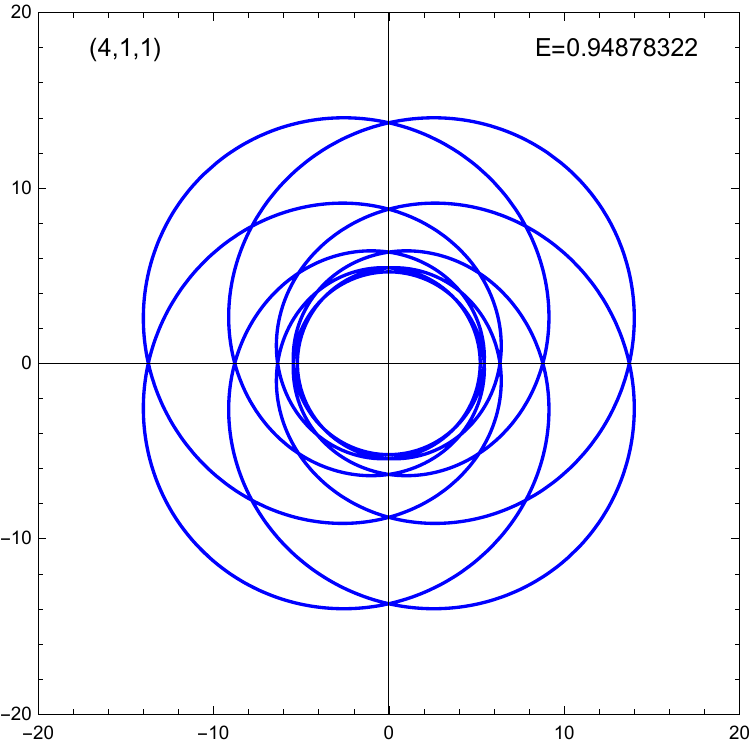} 
\includegraphics[width=5.3cm]{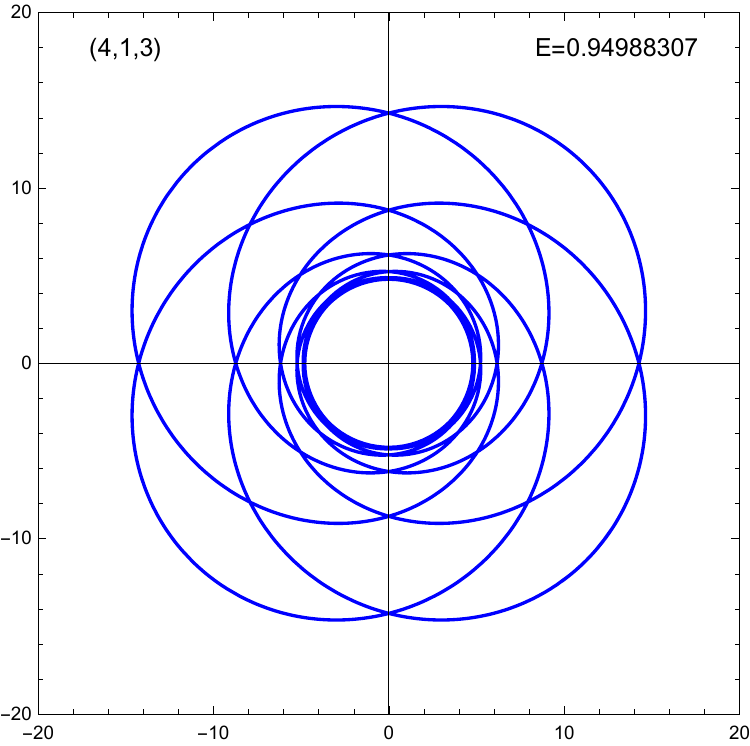} 
\includegraphics[width=5.3cm]{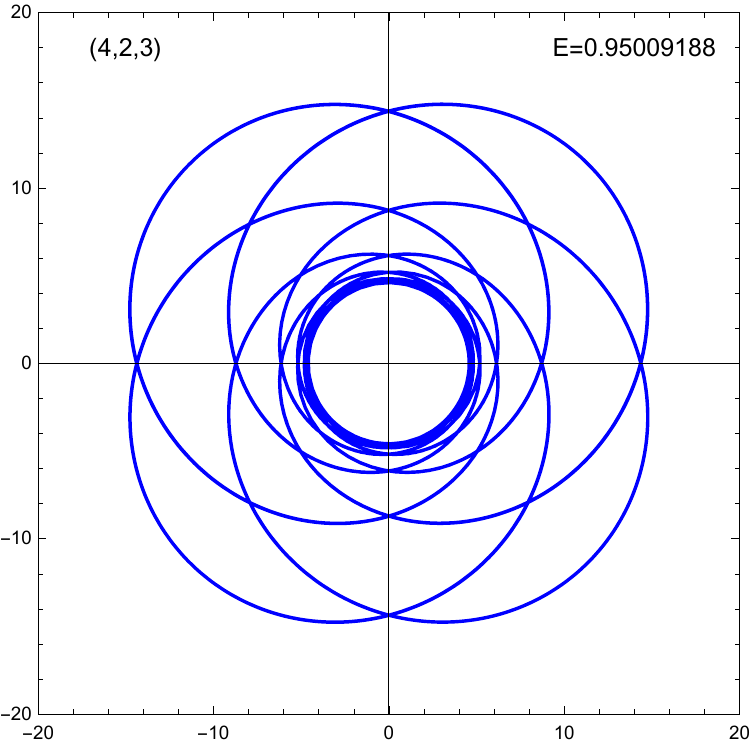} 
\captionsetup{justification=raggedright,singlelinecheck=false}
\caption{Periodic orbits for different values of $(z, w, v)$ around a black hole with DM halo. The value of parameters are $M=10$, $a_0=1000$ and $\epsilon=0.3$.}
\label{periodic2}
\end{figure*}

\begin{table*}
  \centering
\captionsetup{justification=raggedright,singlelinecheck=false}
\caption{\label{table1} For $M_H=10$ and $\epsilon=0.3$, the energy $E$ is computed for orbits characterized by different $(z,w,v)$ and length scale parameter $a_0$. The Schwarzschild case corresponds to $M_H=0$.}
\begin{ruledtabular}
\begin{tabular}{cccccccccc} 
$M_H$ & $a_0$ & $E{(1,1,0)}$ & $E{(1,2,0)}$ & $E{(2,1,1)}$ & $E{(2,2,1)}$ & $E{(3,1,2)}$ & $E{(3,2,2)}$ & $E{(4,1,3)}$ & $E{(4,2,3)}$ \\
 \hline
  0 & - & 0.953628 & 0.957086 & 0.956607 & 0.957170 & 0.956864 & 0.957178 & 0.956946 & 0.957181\\
 \hline
 10 & $10^4$ & 0.952931 & 0.956371 & 0.955896 & 0.956454 & 0.956152 & 0.956462 & 0.956232 & 0.956465\\
 \hline
 10 & $10^3$ & 0.946727 & 0.950009 & 0.949570 & 0.950083 & 0.949808 & 0.950089 & 0.949883 & 0.950092 \\
 \hline
10 & $10^2$ & 0.890310 & 0.891945 & 0.891767 & 0.891968 & 0.891868 & 0.891969 & 0.891898 & 0.891970\\
\end{tabular}
\end{ruledtabular}
\end{table*}

\begin{table*}
\captionsetup{justification=raggedright,singlelinecheck=false}
\caption{For $M_H=10$ and $\epsilon=0.5$, the energy $E$ is computed for orbits characterized by different $(z,w,v)$ and length scale parameter $a_0$. The Schwarzschild case corresponds to $M_H=0$.}
\label{table2}
\begin{ruledtabular}
\begin{tabular}{cccccccccc} 
$M_H$ & $a_0$ & $E{(1,1,0)}$ & $E{(1,2,0)}$ & $E{(2,1,1)}$ & $E{(2,2,1)}$ & $E{(3,1,2)}$ & $E{(3,2,2)}$ & $E{(4,1,3)}$ & $E{(4,2,3)}$ \\ 
 \hline
0 & - & 0.965425 & 0.968383 & 0.968026 & 0.968434 & 0.968225 & 0.96885 & 0.968285 & 0.96844\\
 \hline
10 & $10^4$ & 0.964914 & 0.967853 & 0.967500 & 0.967904 & 0.967697 & 0.967908 & 0.967756 & 0.967909\\
 \hline
10 & $10^3$ & 0.960372 & 0.963136 & 0.962817 & 0.963180 & 0.962997 & 0.963183 & 0.963050 & 0.963184 \\
 \hline
10 & $10^2$ & 0.917747 & 0.918741 & 0.918650 & 0.918750 & 0.918703 & 0.918750 & 0.918718 & 0.918751\\
\end{tabular}
\end{ruledtabular}
\end{table*}

Building on the taxonomy established in \cite{Levin:2008ci}, we define the frequency ratio $q$ as the relationship between the radial ($\omega_r$) and azimuthal ($\omega_\phi$) oscillation frequencies. This ratio is expressed in terms of three integers $(z, w, v)$, zoom $z$, whirl $w$ and vertex $v$ as
\begin{equation}\label{qradial}
q = \frac{\omega_\phi}{\omega_r} - 1 = w + \frac{v}{z}.
\end{equation}
The frequency ratio $q$ is also linked to the equatorial angle $\Delta\phi$ traversed during one radial period by the relation:
\begin{equation}
\frac{\omega_\phi}{\omega_r} = \frac{\Delta\phi}{2\pi},
\end{equation}
where $\Delta\phi$ is given by Eq. (\ref{apsidal}). For irrational values of $q$, the particle's trajectory traces a precessing orbit, characterized by a precession angle $w = \Delta\Phi - 2\pi$. Rational values of $q$, however, correspond to periodic orbits, where the particle returns to its initial position after a finite time. As shown in \cite{Levin:2008ci}, generic orbits can be interpreted as perturbations of these periodic orbits. Thus, the study of periodic orbits provides critical insights into the behavior of generic orbits and the gravitational radiation emitted near black holes.

Given that the angular momentum for bound orbits lies between the values of ISCO and MBO, we can express angular momentum, $L$, as
\begin{equation}
L = L_{\rm ISCO} + \epsilon (L_{\rm MBO} - L_{\rm ISCO})
\end{equation}
where $\epsilon$ is a parameter ranging from 0 to 1. Specifically, $\epsilon = 0$ corresponds to the $L_{\rm ISCO}$, and $\epsilon = 1$ corresponds to the $L_{\rm MBO}$. The constraint $0 \le \epsilon \le 1$ ensures that we remain within the regime of bound orbits.

Fig.~\ref{qe} and Fig.~\ref{ql} illustrate the behavior of the rational number $q$ with respect to energy $E$ and angular momentum $L$, respectively, for various black hole masses $M_H$. Fig.~\ref{qe} reveals a gradual increase in $q$ with increasing $E$, while it reaches a limited value as energy reaches its maximum. This maximum energy exhibits an inverse relationship with $M_H$, decreasing as $M_H$ increases. Conversely, Fig.~\ref{ql} demonstrates a gradual decrease in $q$ with increasing angular momentum $L$, ceasing at the minimum value of $L$. Furthermore, both figures indicate a direct relationship between $L$ and $E$ for a fixed $q$: as $E$ increases, $L$ also increases. In general, the range of $q$ depends on the value of angular momentum $L$. As we mentioned above, when one fixed $L$, $q$ reaches its maximum value as energy reaches its maximum \cite{Levin:2008mq}. It is interesting to note that when $L=L_{\rm MBO}$, the orbit with maximum energy corresponds to the homoclinic orbit \cite{Levin:2008mq}, which has divergent number of whirls, indicating $q\to+\infty$. 

To illustrate the periodic orbits and their GW radiations, we focus on certain periodic orbits with relatively small values of $q$ for two main reasons. First, high $q$ orbits correspond to extreme ``whirl-dominated" motion near ISCO. These orbits in general are short-lived due to rapid gravitational wave radiation, thus are not expected to be astrophysical relevant. Second, orbits with large $q$ require finely tuned initial conditions and are highly sensitive to perturbations, such as those caused by gravitational wave back-reaction. This sensitivity is evident from Fig.~\ref{qe} and Fig.~\ref{ql}, where small changes in energy or angular momentum lead to rapid increases in $ q $. Then by fixing energy to $E=0.96$ in Fig.~\ref{periodic1} and parameter $\epsilon=0.3$ in Fig.~\ref{periodic2}, periodic orbits are illustrated using different combinations of integers $(z, w, v)$. The value of $z$ determines the number of blades in the orbit's shape. Larger $z$ values correspond to larger blade profiles and increasingly complex trajectories.

Table.~\ref{table1} and Table.~\ref{table2} list the values of the frequency ratio $q$ and energy $E$ for periodic orbits with $\epsilon=0.3$ and $\epsilon=0.5$, respectively. These results demonstrate that the energy of the test particle decrease as the length scale $a_0$ varies, keeping $M_H$ fixed. A comparison reveals that periodic orbits around a black hole surrounded by DM exhibit lower energy levels than those around a classical Schwarzschild black hole.

\section{Gravitational Waveform from Periodic Orbits}\label{GW}
\renewcommand{\theequation}{5.\arabic{equation}} \setcounter{equation}{0}

EMRIs are promising sources for future space-based GW detectors. These systems, comprising a stellar-mass compact object orbiting a SMBH, emit GWs that encode rich information about the system's dynamics and the surrounding spacetime.  When the smaller object is in a periodic orbit around a SMBH immersed in a DM halo, the emitted GW waveform offers a unique opportunity to study distinct features that can be exploited to probe the system's properties. This section outlines the theoretical framework for calculating gravitational waveforms from such periodic orbits.

The computation of gravitational waveforms from EMRIs often relies on the adiabatic approximation, which is valid when the inspiral timescale is much longer than the orbital period. Under this approximation, the smaller object's energy and angular momentum can be considered constant over several orbits, allowing the trajectory to be modeled as a geodesic in the background spacetime of the SMBH. The back reaction of gravitational radiation on the motion of the smaller object is effectively neglected \cite{Hughes:1999bq,Isoyama:2021jjd}. 

\begin{figure*}
\includegraphics[width=12.8cm]{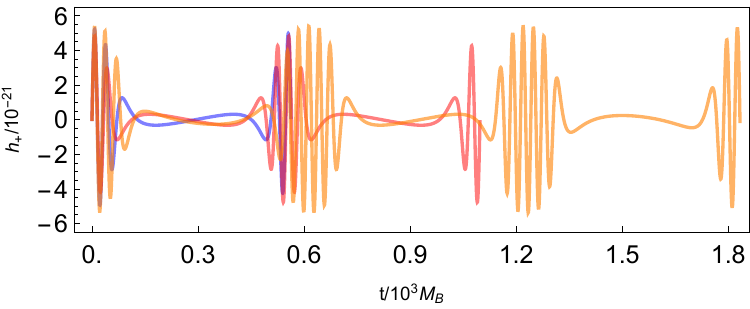}
\includegraphics[width=12.8cm]{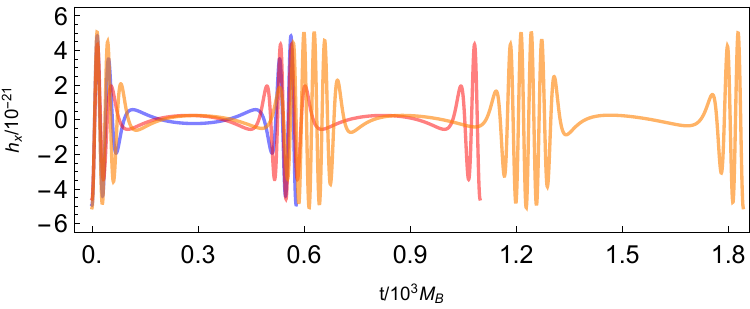} 
\captionsetup{justification=raggedright,singlelinecheck=false}
\caption{Gravitational waveforms from a test object with $m=10 M_\odot$ around periodic orbits (1,2,0): blue, (2,1,1): red, (3,2,2): orange, around a SMBH immersed in a DM halo with $M_B= 10^6  M_\odot$. Other parameters are  $M_H=10$ and $a_0=1000$. The energy is fixed at $E=0.96$.}
\label{wave1}
\end{figure*}

A numerical ``kludge" waveform model provides a practical approach to calculating the GWs from these periodic orbits within the DM halo environment \cite{Babak:2006uv}. This method involves a two-step process: first, the orbit of the small object is determined by numerically solving the geodesic equations of motion in the black hole's spacetime, accounting for the DM distribution. Subsequently, the quadrupole formula for gravitational radiation is applied to this calculated orbit, generating the corresponding waveform. This approach allows for a preliminary investigation of the GW signals emitted by EMRIs and their potential to reveal insights into the properties of both the orbit and the central black hole, as well as the distribution of DM surrounding it.

\begin{figure*}
  \centering
  \begin{tabular}{ c }
    \includegraphics[width=0.4\textwidth]{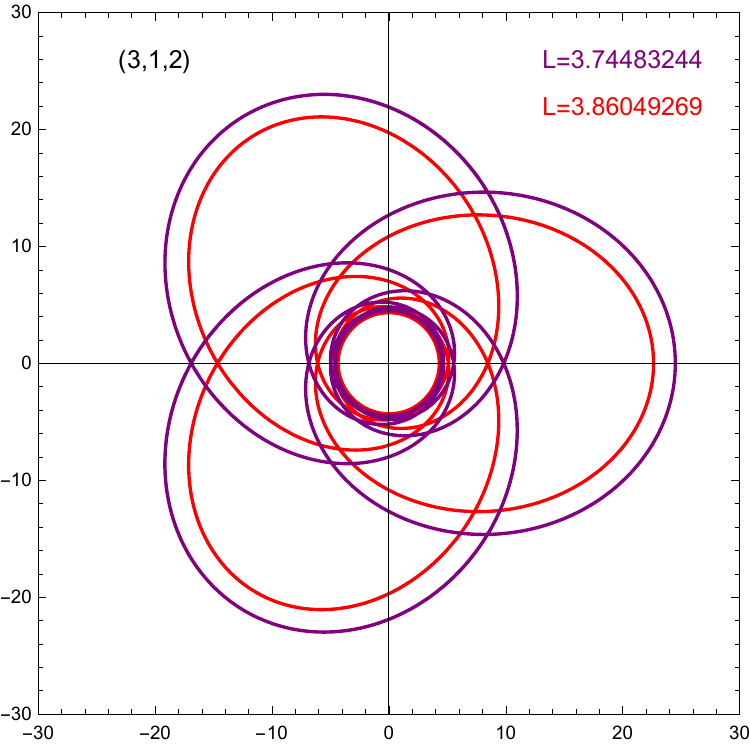}
  \end{tabular}%
  \begin{tabular}{ c c }
   \includegraphics[width=0.5\textwidth]{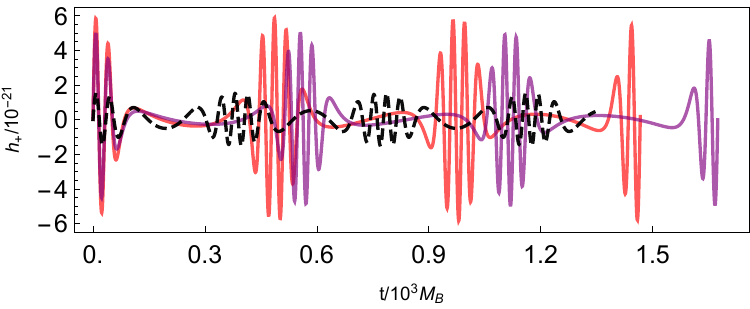}\\ 
    \includegraphics[width=0.5\textwidth]{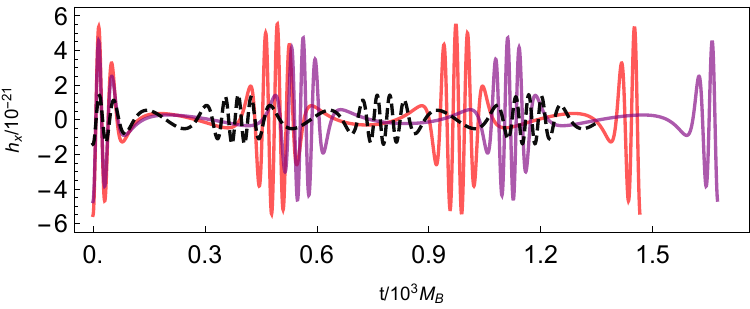} 
     \end{tabular}
\captionsetup{justification=raggedright,singlelinecheck=false}
\caption{The left figure is a sketch figure which shows a typical periodic orbit around a black hole with $(z,w, v) = (3, 1, 2)$. In the right figure, the dotted line represents the GW of the Schwarzschild black hole, while the purple solid line represents the GW of a black hole immersed in DM halo when $M_H=10 M_B$ and $a_0=1000 M_B$. The red solid line represents the GW of a black hole immersed in DM halo when $M_H=10$ and $a_0=100$.}
\label{wave2}
\end{figure*}

\begin{figure*}
  \centering
  \begin{tabular}{ c }
    \includegraphics[width=0.4\textwidth]{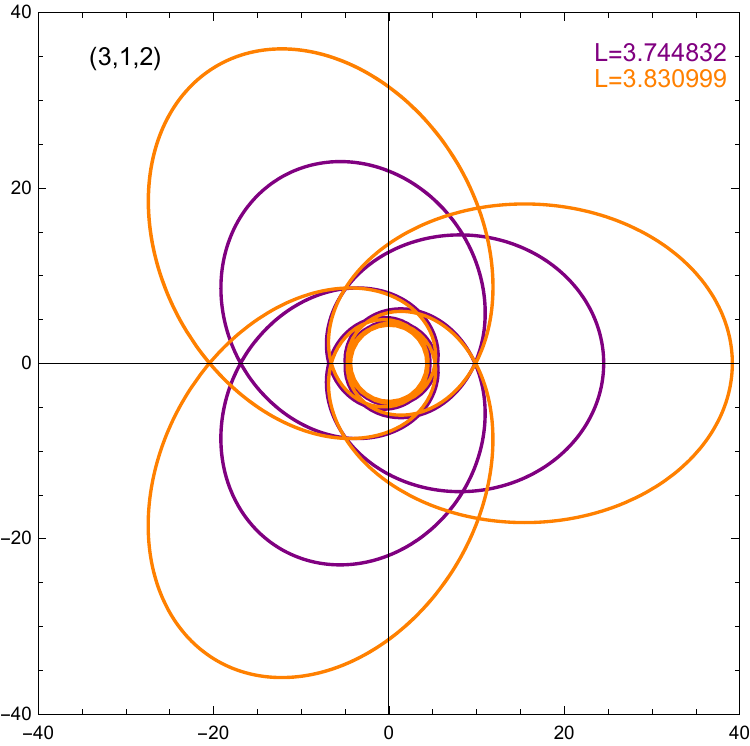}
  \end{tabular}%
  \begin{tabular}{ c c }
   \includegraphics[width=0.5\textwidth]{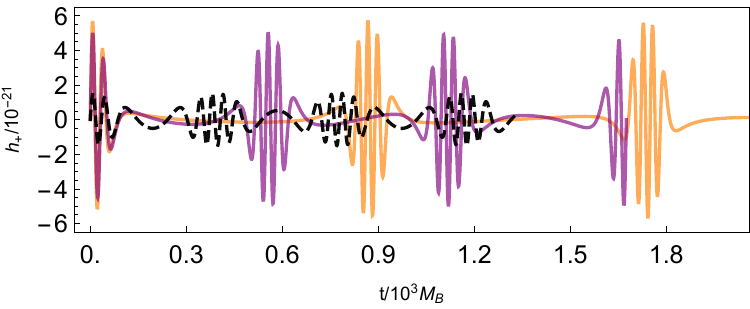}\\ 
    \includegraphics[width=0.5\textwidth]{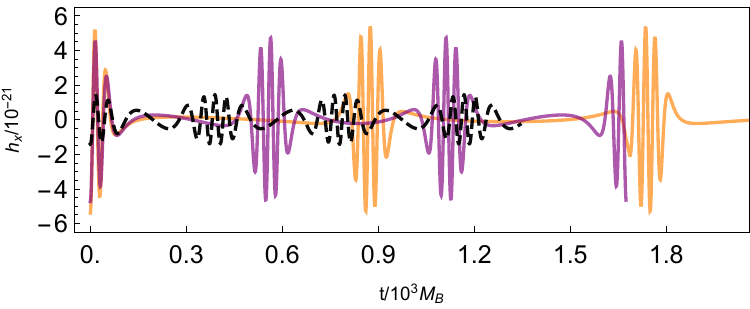} 
     \end{tabular}
 \captionsetup{justification=raggedright,singlelinecheck=false}
\caption{The left figure is a sketch figure which shows a typical periodic orbit around a black hole with $(z,w, v) = (3, 1, 2)$. In the right figure, the dotted line represents the GW of the Schwarzschild black hole, while the purple solid line represents the GW of a black hole immersed in DM halo when $M_H=10 M_B$ and $a_0=1000 M_B$. The orange solid line represents the GW of a black hole immersed in DM halo when $M_H=20 M_B$ and $a_0=1000 M_B$. }
\label{wave3}
\end{figure*}

For a metric perturbation $h_{ij}$ representing the GW and the symmetric and trace-free (STF) mass quadrupole $I_{ij}$, the quadrupole formula for gravitational radiation is given by
\begin{equation}
  h_{ij}=\frac{1}{A}\ddot{I}_{ij},
\end{equation}
where $A=c^4 D_L/(2G)$, $G=1=c$, $D_L$ is the luminosity distance to the source. By numerically solving the geodesic equations of motion, the trajectory $Z_i(t)$ of the smaller object in the curved spacetime of the SMBH immersed in DM halo can be determined.  This trajectory is crucial for the subsequent calculation of the gravitational waveform. For a small object of mass $m$ following a trajectory $Z^i(t)$, the $I_{ij}$ is given by \cite{Thorne:1980ru}
\begin{equation}\label{lvalue}
I^{ij}=m\int d^3 x^i x^j  \delta^3(x^i - Z^i(t)).
\end{equation}

\begin{figure*}
\includegraphics[width=8.cm]{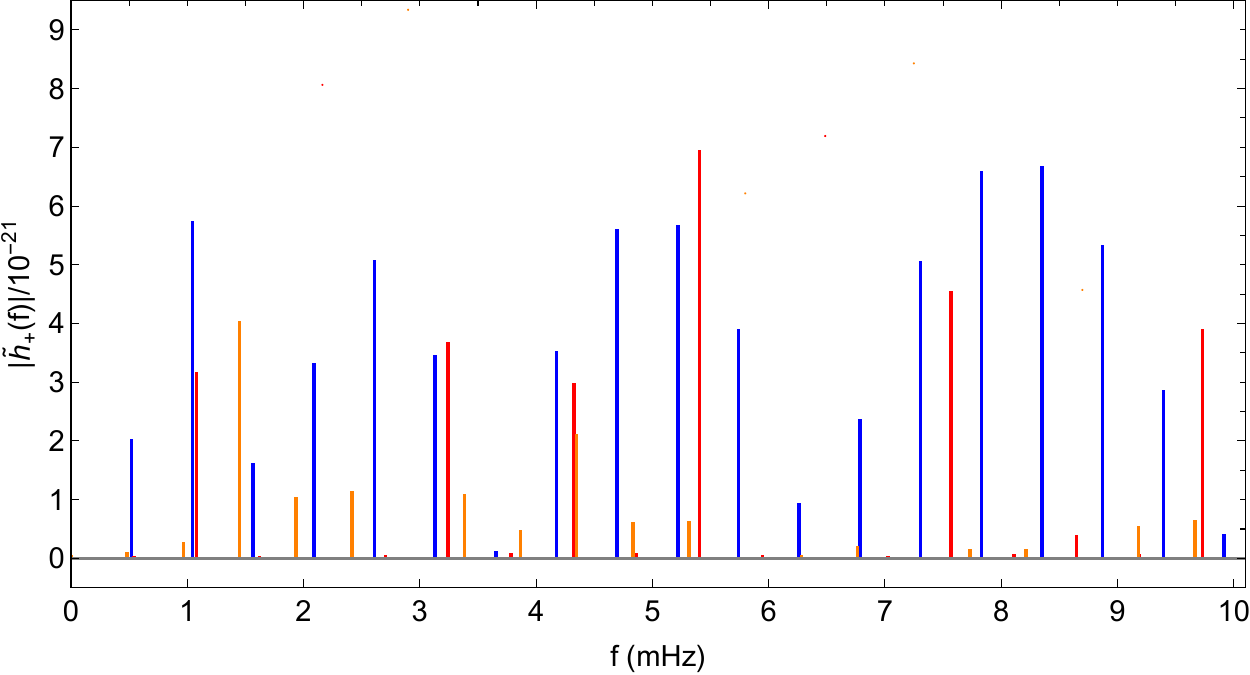}
\includegraphics[width=8.cm]{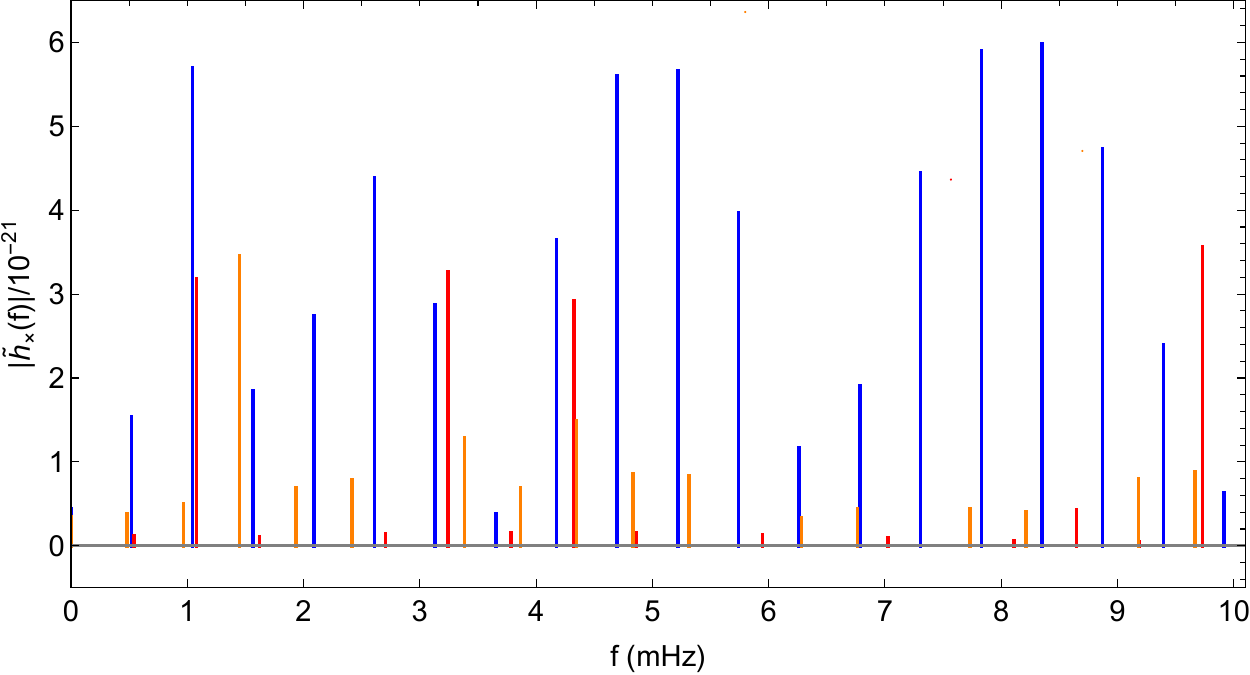} 
\captionsetup{justification=raggedright,singlelinecheck=false}
\caption{Absolute frequency spectra $\vert \tilde{h}_{+, \times (f)} \vert$ corresponding to the waveforms in Figure \ref{wave1}.}
\label{FS1}
\end{figure*}

\begin{figure*}
\includegraphics[width=8.cm]{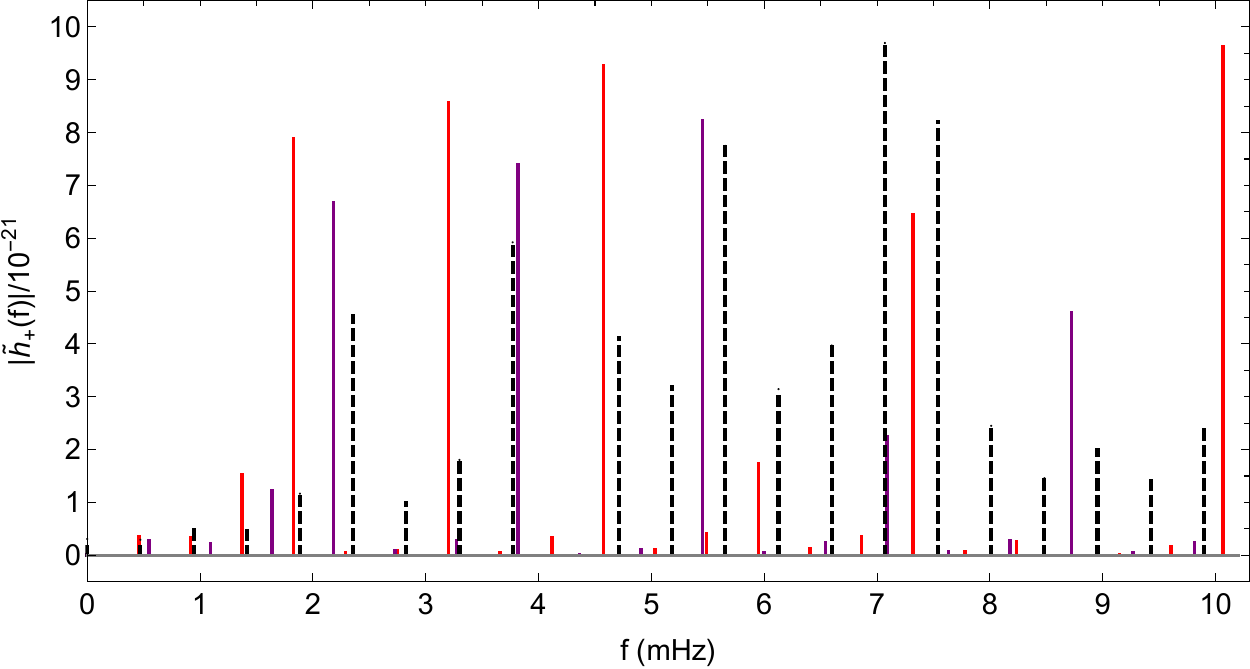}
\includegraphics[width=8.cm]{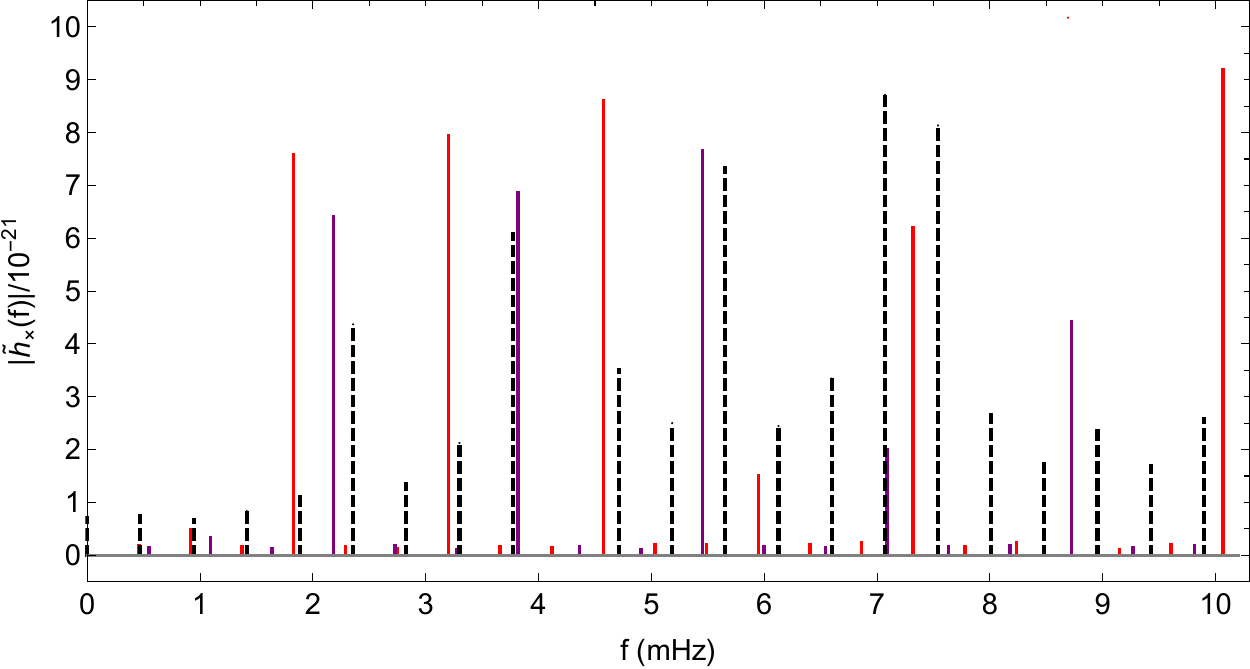} 
\captionsetup{justification=raggedright,singlelinecheck=false}
\caption{Absolute frequency spectra $\vert \tilde{h}_{+, \times (f)} \vert$ corresponding to the waveforms in Figure \ref{wave2}.}
\label{FS2}
\end{figure*}

\begin{figure*}
\includegraphics[width=8.cm]{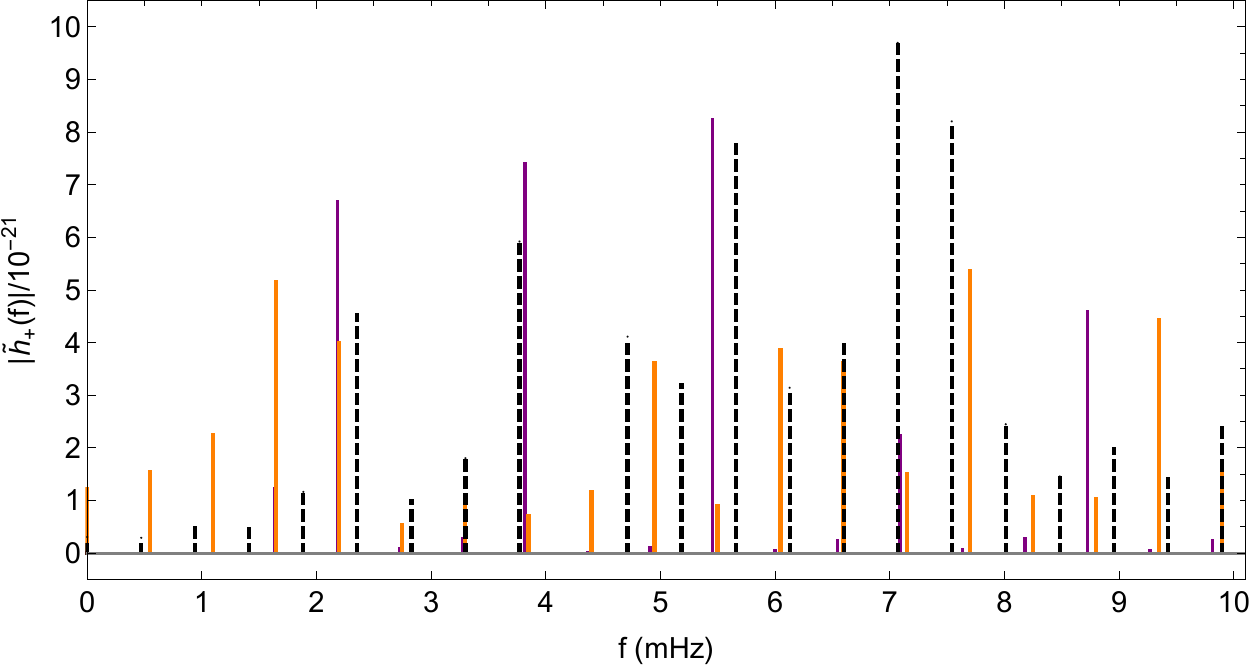}
\includegraphics[width=8.cm]{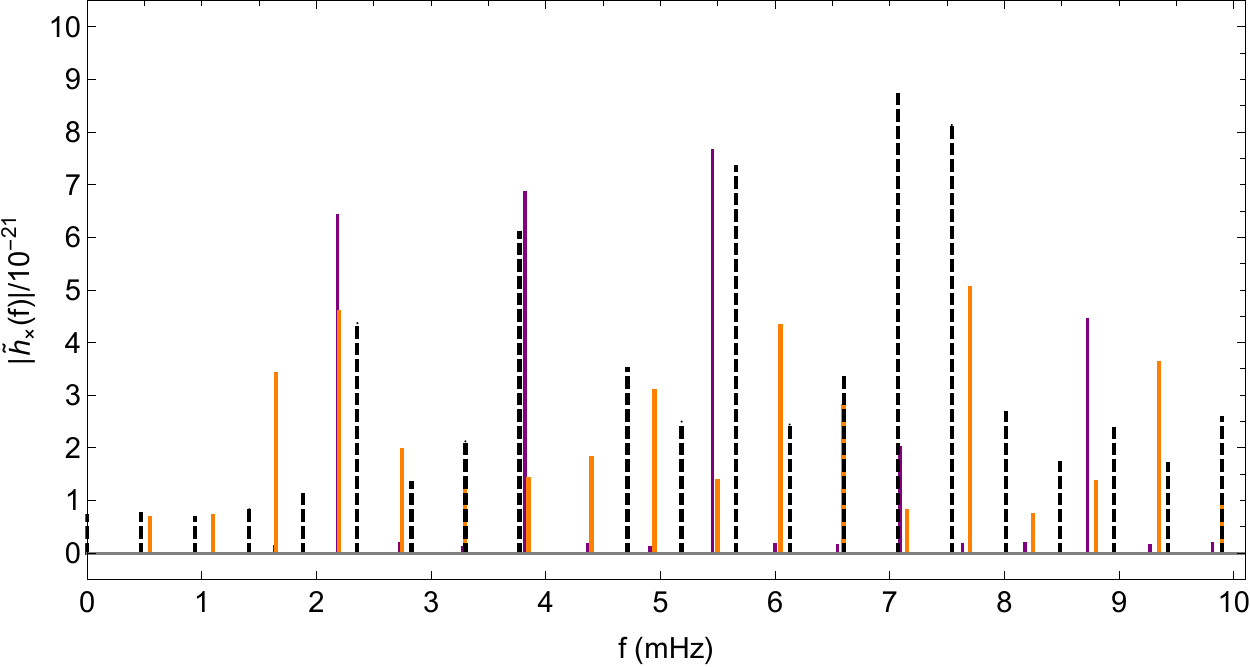} 
\captionsetup{justification=raggedright,singlelinecheck=false}
\caption{Absolute frequency spectra $\vert \tilde{h}_{+, \times (f)} \vert$ corresponding to the waveforms in Figure \ref{wave3}.}
\label{FS3}
\end{figure*}

 The choice of coordinate system plays a crucial role in the calculation and interpretation of gravitational waveforms. While the geodesic equations are often solved in Boyer-Lindquist coordinates $(r, \theta, \phi)$, the waveform is typically expressed in a detector-adapted coordinate system $(X, Y, Z)$.  This transformation simplifies the analysis of the signal as measured by a GW detector.  The transformation from Boyer-Lindquist to Cartesian coordinates is given by \cite{Babak:2006uv}
\begin{equation}\label{4.3}
x=r \sin\theta \cos\phi, \quad y=r\sin\theta\sin\phi,    z=r \cos\theta.
\end{equation}
This allows us to project the small object's trajectory onto a Cartesian grid.
The metric perturbations $h_{ij}$, representing the GWs, are then calculated using the second time derivative of the mass quadrupole moment $I_{ij}$ as
\begin{equation}\label{4.4}
h_{ij}=
\frac{m}{A}(a_i x_j+a_j x_i+2 v_i v_j),
\end{equation}
where 
$a_i$ and $v_i$ are its acceleration and velocity components, respectively.

Now, to analyze the GW signal at a detector, a detector-adapted coordinate system $(X, Y, Z)$ is introduced, centered on the black hole and oriented relative to the original $(x, y, z)$ system by the inclination angle $\iota$ and the longitude of pericenter $\zeta$. The unit vectors of the detector frame in the $(x, y, z)$ coordinates are:
\begin{eqnarray}
\hat{e}_X &=& (\cos\zeta, -\sin\zeta, 0),\\
\hat{e}_Y &=& (\sin\iota \sin\zeta, \cos\iota \cos\zeta, -\sin\iota),\\
\hat{e}_Z &=& (\sin\iota \sin\zeta, -\sin\iota \cos\zeta, \cos\iota),
\end{eqnarray}
The GW polarizations $h_+$ and $h_\times$ are then obtained by projecting $h_{ij}$, Eq. (\ref{4.4}), onto the detector frame
\begin{eqnarray}\label{4.5}
h_+&=\frac{1}{2}\big(e_X^i e_X^j-e_Y^ie_Y^j\big)h_{ij},\\
h_{\times}&=\frac{1}{2}\big(e_X^i e_Y^j-e_Y^ie_X^j\big)h_{ij},
\end{eqnarray}
These polarizations can be expressed in terms of components $h_{\zeta\zeta}$, $h_{\iota\iota}$, and $h_{\iota\zeta}$ which are defined in the detector frame as combinations of the $h_{ij}$ components as
\begin{eqnarray}\label{4.64}
h_+&=&\frac{1}{2}\big(h_{\zeta\zeta}-h_{\iota\iota}\big),\\\label{4.6}
h_{\times}&=&h_{\iota\zeta},
\end{eqnarray}
where the components are \cite{Babak:2006uv}
\begin{eqnarray}\label{4.7}
h_{\zeta\zeta}&=&h_{xx}\cos^2\zeta-h_{xy}\sin{2 \zeta}+h_{yy}\sin^2\zeta,\\
h_{\iota\iota}&=& \cos^2\iota\big[h_{xx}\sin^2\zeta + h_{xy}\sin 2 \zeta + h_{yy}\cos^2 \zeta\big] \nonumber \\
&& +h_{zz} \sin^2\iota - \sin{2 \iota}\big[h_{xz \sin\zeta}+h_{yz}\cos\zeta\big],\\
h_{\iota\zeta}&=& \frac{1}{2}\cos\iota\big[h_{xx} \sin {2 \zeta}+ 2 h_{xy}\cos{2 \zeta}- h_{yy}\sin{2 \zeta}\big]\nonumber\\
&&+\sin\iota\big[h_{yz} \sin\zeta-h_{xx} \cos\zeta\big].
\end{eqnarray}
To demonstrate the impact of DM halo effects on gravitational waveforms from various periodic orbits in an extreme-mass-ratio inspiral system, we analyze a scenario with a small object of mass $m = 10 M_\odot$ orbiting a SMBH of mass $M = 10^6 M_\odot$. For simplicity, the inclination angle $\iota$ and the latitudinal angle $\zeta$ are set to $\pi/4$, and a luminosity distance $D_L$ of $2$ Gpc is assumed to calculate the GW's polarizations. 

In the GW signals labeled $h_+$ and $h_\times$, there is a clear pattern of alternating behavior. During the parts of the orbit where the path stretches out into an elongated ellipse (the zoom phases), the signal remains subdued. This calm period is then followed by brief, intense bursts when the orbit becomes nearly circular (the whirl phases). Interestingly, the number of these calm intervals matches the number of distinct segments in the orbit, while the number of intense bursts aligns with the number of circular loops. The numerical results for Eqs.~(\ref{4.64}) and (\ref{4.4}) are computed and displayed in Fig.~\ref{wave1}, Fig.~\ref{wave2} and Fig.~\ref{wave3}. These figures illustrate the distinct ``zoom" and ``whirl" phases present in gravitational waveforms from periodic orbits in EMRIs, reflecting the orbital motion of the smaller object, in one complete period.

In Fig.~\ref{wave1}, the gravitational waveforms are displayed while increasing the zoom number $z$ as $(z,w,v)=(1,2,0),(2,1,1)$ and $ (3,2,2)$. This analysis demonstrates a strong link between gravitational waveforms and the small object's orbital motion. Each orbit displays clear ``zoom" and ``whirl" phases in the waveform that mirror the corresponding behaviors in the object's trajectory. Additionally, orbits with higher zoom numbers $z$ produce waveforms with more intricate substructures, reflecting the increased number of ``leaves" in the full periodic orbit.

The presence of DM halo considerably affects the gravitational waveform produced by the massive particle moving in the periodic orbit. We again take periodic orbit $(z,w,v)=(3,1,2)$ as shown in Fig.~\ref{wave2} and Fig.~\ref{wave3}. Our finding reveals that when the DM halo's scale length  $a_0$ is reduced or its mass $M_H$ is increased, the gravitational waveforms experience a noticeable shift in phase along with a significant boost in amplitude. In comparison to the Schwarzschild case, the phase alteration is considerable at $a_0 = 100 M_B$ and becomes even more pronounced at $a_0 = 1000 M_B$, accompanied by drastic changes in amplitude. Similarly,  variations in $M_H$ affect both the phase and amplitude of the GWs. 

The gravitational waveforms emitted by a test particle moving along periodic orbits around a supermassive black hole in a dark matter halo can be further analyzed through their frequency spectra $\vert \tilde{h}_{+,\times}\vert$ and characteristic strain $h_c(f)$, defined as
\be\label{ch}
h_c(f)= 2 f \left(\vert \tilde{h}_+(f) \vert ^2 +\vert \tilde{h}_\times(f) \vert ^2\right)^{1/2},
\ee
To obtain the frequency spectra, we apply a discrete Fourier transform (DFT) to the time-domain gravitational waveforms, converting the signal from the time domain to the frequency domain. This allows for a detailed examination of the frequency distribution present in the signal, revealing the influence of the periodic orbital motion of the particle on the frequency structure of the gravitational wave signal, see Figs.~\ref{FS1}, \ref{FS2} and \ref{FS3}.  The characteristic frequencies of gravitational waves from these periodic orbits fall primarily within the millihertz range, which is particularly significant for space-based gravitational wave detectors such as LISA, optimized for detecting low-frequency gravitational waves. The characteristic frequencies for different periodic orbits, indexed by $(z,w,v)$, are shown in Fig. \ref{FS1}, while the characteristic frequencies for various values of the parameters $a_0$ and $M_H$ are depicted in Figs. \ref{FS2} and \ref{FS3}, respectively. 
\begin{figure*}
\includegraphics[width=8.cm]{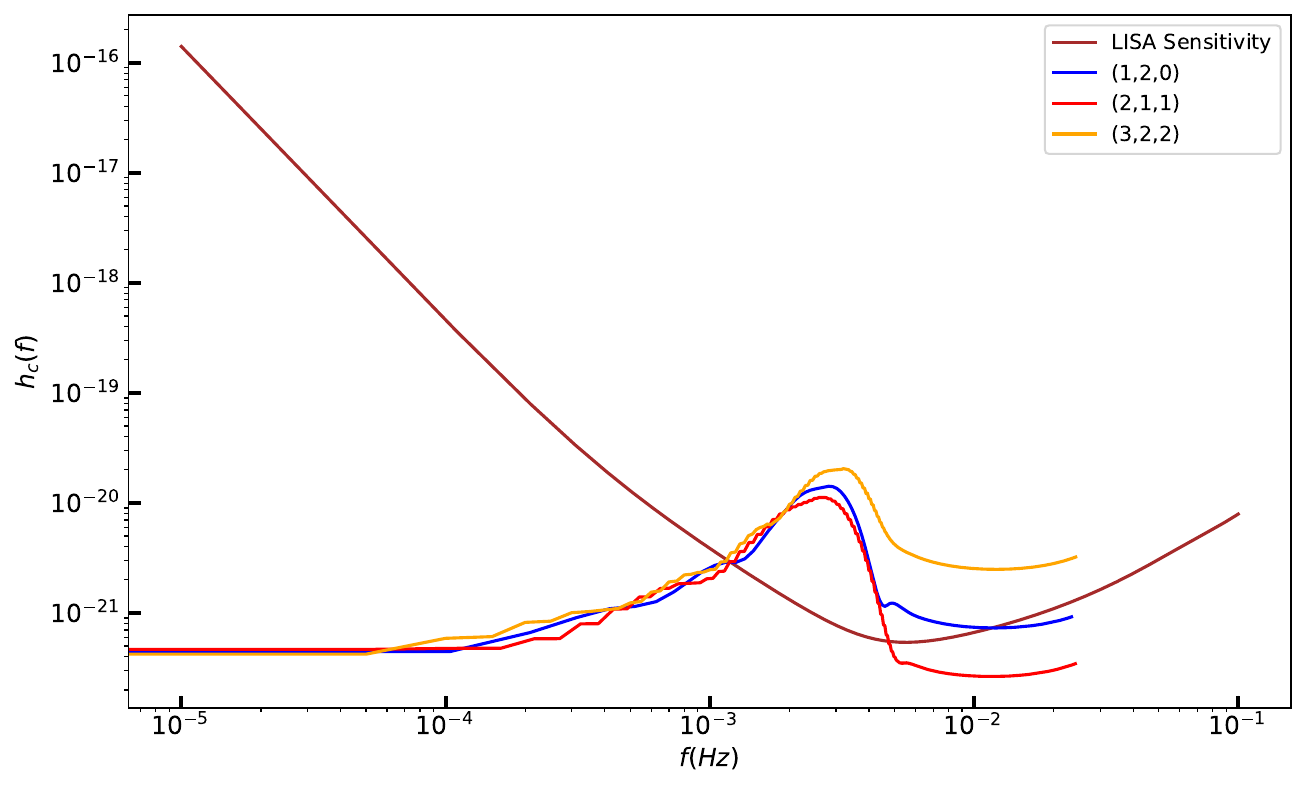}
\includegraphics[width=8.cm]{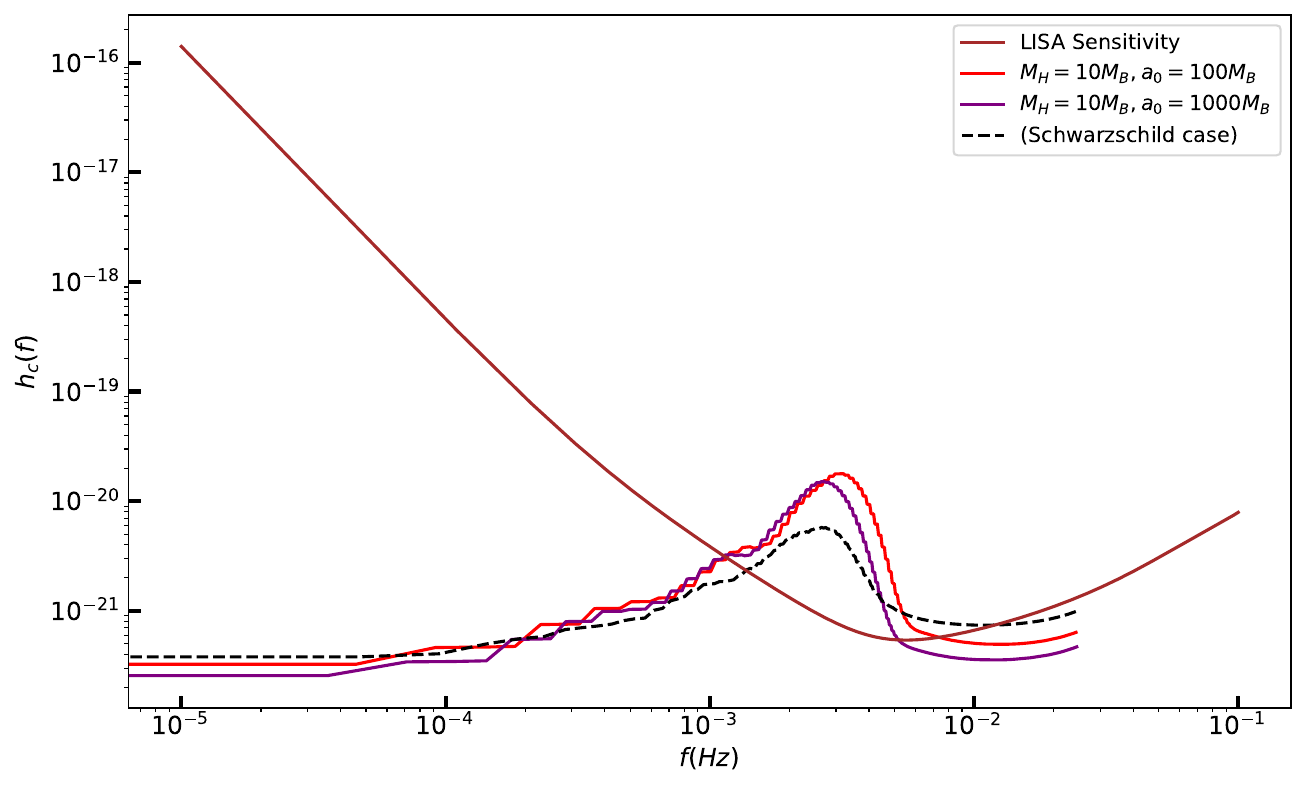} 
\includegraphics[width=8.cm]{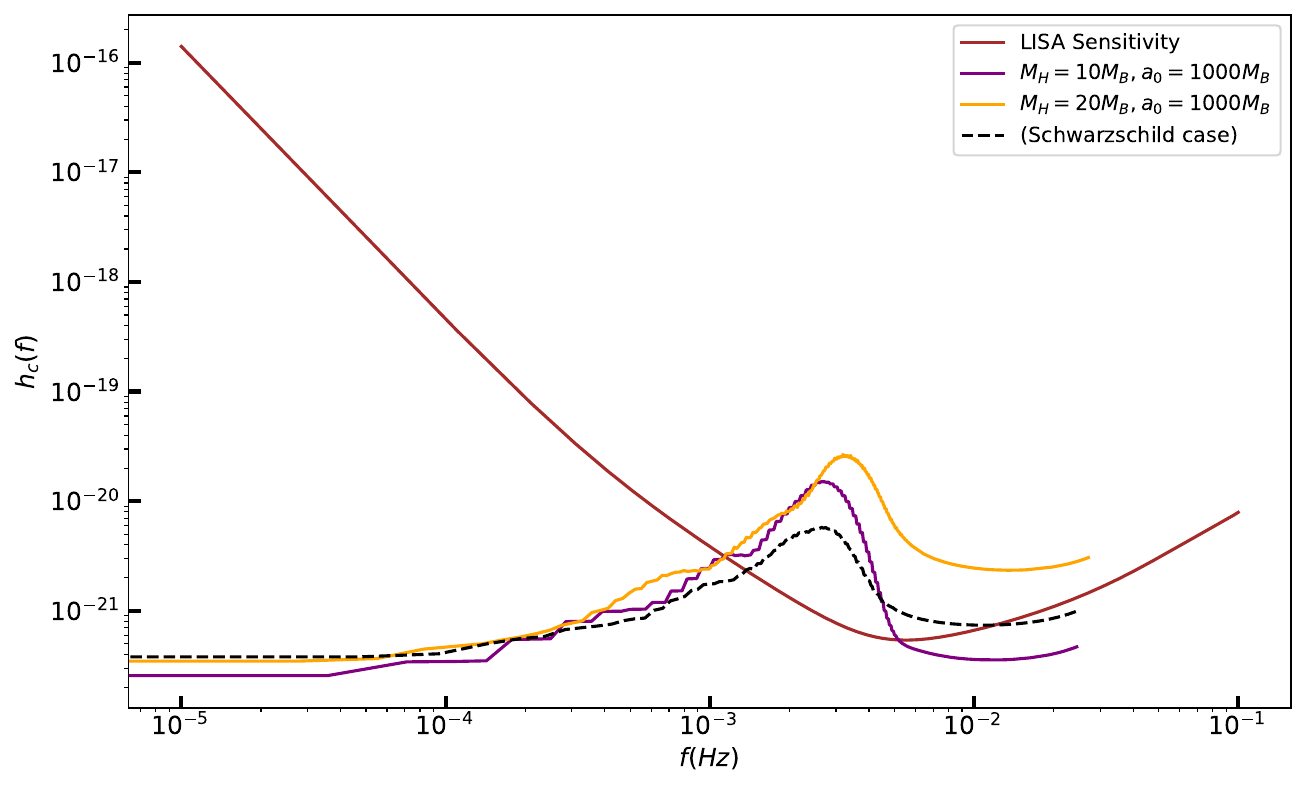} 
\captionsetup{justification=raggedright,singlelinecheck=false}
\caption{Characteristic strain $h_c(f)$ of gravitational waveforms of periodic orbits in Figures  \ref{wave1}, \ref{wave2} and  \ref{wave3}. }
\label{CS}
\end{figure*}
To enhance the visual clarity of the plots for characteristic strain (\ref{ch}), we applied a smoothing technique to the numerically generated $h_c(f)$ by performing a running average over 30 frequency bins. Selecting a larger window size would have smoothed out more of the noise but may have blurred some of the finer features in the data. It can be observed, in Fig. \ref{CS}, that some segments of the characteristic strain, corresponding to different orbital configurations $(z,w,v)$ and parameters $a_0$ and $M_H$, exceed the sensitivity curve of LISA (Laser Interferometer Space Antenna). This suggests that these gravitational waves, characterized by their distinctive zoom-whirl features in the presence of a dark matter halo, are likely to be detectable by LISA, offering valuable insights into the spacetime geometry surrounding supermassive black holes.

\section{Conclusion and Outlook}\label{conclusion}
\renewcommand{\theequation}{7.\arabic{equation}} \setcounter{equation}{0}

This study explored the impact of the DM halo on the orbits of a particle around a black hole. Unlike the classic Schwarzschild case, the presence of the halo significantly altered these orbits. By numerically analyzing the effective potential, the study determined the characteristics of MBOs and ISCOs. Increasing the halo's characteristic length scale $a_0$ increases $r_{\rm MBO}$ but reduces their $L_{\rm MBO}$. Conversely, increasing the halo mass ($M_H$) decreased $r_{\rm MBO}$ while increased $L_{\rm MBO}$. Similar trends emerged for ISCOs: with fixed $M_H$, increasing $a_0$ led to increase in $r_{\rm ISCO}$, higher $E_{\rm ISCO}$, but lower $L_{\rm ISCO}$. With fixed $a_0$, increasing $M_H$ had the opposite effect on $r_{\rm ISCO}$ and $E_{\rm ISCO}$ but increased $L_{\rm ISCO}$. The study also analyzed the allowed parameter region in $(L, E)$ by mapping the energy $E$ and angular momentum $L$ combinations permitting bound orbits, illustrating the overall influence of the DM halo on particle motion around the black hole.

Next, we studied the influence of the parameters $M_H$ and $a_0$ on the periodic orbit of a massive particle around the black hole in the DM halo. The relationship between the rational number, $q$, and both the energy $E$ and angular momentum $L$ of these orbits was numerically analyzed. Results indicate that for a fixed $q$, $E$ decreases with increasing $a_0$ (for constant $M_H$). Similarly, for a fixed $q$, $L$ also exhibits a dependence on $a_0$ (for constant $M_H$). The maximum $E$ and minimum $L$ for these periodic orbits were also found to be influenced by the DM halo parameters. Specifically, the maximum $E$ decreases with increasing $M_H$, while the minimum $L$ demonstrates a dependence on $M_H$. Overall, the presence of the DM halo significantly alters the characteristics of the periodic orbits compared to a standard Schwarzschild black hole, as evidenced by the lower energy levels observed in the presence of the halo.

Finally, we analyze an EMRI system consisting of a test object with mass $ m = 10 M_\odot $ following periodic orbits around a SMBH, having mass $M_B = 10^6 M_\odot$, immersed in DM halo. Using the numerical kludge scheme, we investigated the resulting gravitational waveforms by positioning the system at a luminosity distance of $D_L = 2$ Gpc from the detector, with an inclination angle of $\iota = \pi/4$ and a longitude of pericenter  $\zeta = \pi/4 $. This study demonstrates a clear correlation between the gravitational waveforms emitted by a small object orbiting a SMBH and the object's zoom-whirl orbital behavior. Higher zoom numbers correspond to more complex waveform substructures. Furthermore, the presence of a DM halo significantly impacts these waveforms. Decreasing the halo's scale length $a_0$ or increasing its mass $M_H$ results in both a phase shift and amplitude increase in the waveforms. To assess how well gravitational waves from EMRIs with periodic orbits can be detected, we analyzed their time-domain waveforms using discrete Fourier transforms to extract the frequency spectra. The results indicate that the frequencies of these gravitational waves generally fall within the sensitivity range of space-based detectors. From the spectra, we determined the characteristic strains and observed that, for certain combinations of $(z,w,v)$ or specific values of the parameters $a_0$ and $M$, the strains surpass LISA's sensitivity threshold. This suggests that space-based gravitational wave observatories could capture signals from EMRIs with periodic orbits, providing a promising avenue to explore supermassive black holes with dark matter halos.

The DM distribution alters the orbital dynamics, leading to deviations from the waveforms predicted by GR in a vacuum. 
 In conclusion, our study highlights that the properties of the DM halo play a critical role in shaping GW signals, offering promising potential for future observations to probe the influence of DM in strong gravitational fields.

Here we would like to make a few remarks about the limitation of the calculations of the waveforms and the potential extensions of the current study. First, we use the adiabatic approximation with which we ignore the backreaction of the gravitational radiation to the periodic orbits, which is a valid approximation when considering only a few orbital periods, as is the case in this study. Exploring the impact of gravitational radiation on the long-term evolution of periodic orbits represents an interesting avenue for future research. Second, we also ignores the contributions of multipoles higher than the quadratic order. It is also crucial for developing more accurate waveform models to include higher-order multipole moments in the gravitational wave expansion. At last, once these accurate waveforms become available, we will be able to investigate how future gravitational wave detectors might constrain or test the effects of dark matter on periodic orbits. We look forward to addressing these challenges in subsequent studies.

\section*{Acknowledgements}

This work is supported by the National Key Research and Development Program of China under Grant No. 2020YFC2201503, the National Natural Science Foundation of China under Grants No.~12275238 and No.~11675143, the Zhejiang Provincial Natural Science Foundation of China under Grants No.~LR21A050001 and No.~LY20A050002, and the Fundamental Research Funds for the Provincial Universities of Zhejiang in China under Grant No.~RF-A2019015.

\end{document}